\documentclass[11pt,a4paper]{article}

\usepackage{amssymb}
\usepackage[dvips]{graphicx}
\usepackage{bm}

\begin{document}

\title{The Adler $D$-function for ${\cal N}=1$ SQCD regularized by higher covariant derivatives in the three-loop approximation}

\author{A.L.Kataev\\
{\small{\em Institute for Nuclear Research of the Russian Academy of Sciences,}}\\
{\small{\em 117312, Moscow, Russia}}\\
{\small{\em Moscow Institute of Physics and Technology,}}\\
{\small{\em 141700 Dolgoprudnyi, Russia}}\\
\\
A.E.Kazantsev, K.V.Stepanyantz\\
{\small{\em Moscow State University}}, {\small{\em  Physical
Faculty, Department  of Theoretical Physics}}\\
{\small{\em 119991, Moscow, Russia}}}

\maketitle

\begin{abstract}
We calculate the Adler $D$-function for ${\cal N}=1$ SQCD in the three-loop approximation using the higher covariant derivative regularization and the NSVZ-like subtraction scheme. The recently formulated  all-order relation between the Adler function and the anomalous dimension of the matter superfields defined in terms of the bare coupling constant is first considered and generalized to the case of an arbitrary representation for the chiral matter superfields. The correctness of this all-order relation is explicitly verified at the three-loop level. The special renormalization scheme in which this all-order relation remains valid for the $D$-function and the anomalous dimension defined in terms of the renormalized coupling constant is constructed in the case of using the higher derivative regularization. The analytic expression for the Adler function for ${\cal N}=1$ SQCD is found in this scheme to the order $O(\alpha_s^2)$. The problem of scheme-dependence of the $D$-function and the NSVZ-like equation is briefly discussed.
\end{abstract}

\vspace*{-17.4cm}

\begin{flushright}
INR-TH-2017-017
\end{flushright}

\vspace*{16.5cm}

\section{Introduction}
\hspace*{\parindent}

The Adler $D$-function \cite{Adler:1974gd} is closely related to the normalized cross-section of electron-positron annihilation to hadrons, namely to the ratio

\begin{equation}
\label{ratio}
R(s)=\frac{\sigma(e^+e^-\rightarrow {hadrons})}{\sigma_0(e^+e^-\rightarrow
\mu^+\mu^-)}=12\pi \mbox{Im}\, \Pi(s),
\end{equation}

\noindent
where $\sigma_0(e^+e^-\rightarrow \mu^+\mu^-)=4\pi\alpha_{*}^2/3s$ with $\alpha_{*}$ being the QED fine-structure constant. $\Pi(s)$ is the perturbatively evaluated expression for the hadronic vacuum polarization function, which plays an important role in studies of contributions of the strong interaction to various physical quantities. In particular, it is needed for determining strong interaction contributions to the theoretical expression for the muon anomalous magnetic moment (for the most recent discussion see \cite{Jegerlehner:2017lbd}). The $D$-function is connected with $R(s)$ by the dispersion relation

\begin{equation}
D(Q^2)= - 12\pi^2Q^2\frac{d}{dQ^2}\Pi(Q^2)=Q^2\int_0^{\infty}ds \frac{R(s)}{(s+Q^2)^2}
\end{equation}

\noindent
and can be used to compare the theoretical QCD predictions with the available experimental data for $R(s)$ \cite{Eidelman:1998vc}. In the region where perturbation theory is applicable a theoretical expression for the $D$-function is defined by a sum of diagrams with two external lines of the Abelian gauge field, whereas the quarks and gluons are propagating in the internal loops.

Massless perturbative theoretical expressions for the $D$-function are known in several gauge models. In QCD it was analytically evaluated at the three-loop level (to the order $O(\alpha_s^2)$) in Refs. \cite{Chetyrkin:1979bj,Celmaster:1979xr} and numerically in Ref. \cite{Dine:1979qh}. At the level $O(\alpha_s^3)$ the $D$-function has been analytically calculated in \cite{Gorishnii:1990vf}. This result was confirmed in Refs. \cite{Surguladze:1990tg,Chetyrkin:1996ez}. At present, the QCD analytical expression for the $D$-function is known to the order $O(\alpha_s^4)$ \cite{Baikov:2008jh,Baikov:2012zn}. In Ref. \cite{Chetyrkin:1983qc} the $\alpha_s^2$ corrections to the $D$-function were analytically evaluated in a theoretical model of strong interactions which contains QCD supplemented with multiplets of coloured scalar fields.

A more consistent theoretical model of strong interactions containing coloured scalar fields is ${\cal N}=1$ SQCD, which is continuing to attract attention of both theoreticians and experimentalists (for a review see, e.g., \cite{Mihaila:2013wma}). In this model the $O(\alpha_s)$-corrections to the $D$-function have been evaluated in \cite{Kataev:1983at,Altarelli:1983pr}.

An all-loop exact formula relating the $D$-function of ${\cal N}=1$ SQCD to the anomalous dimension of the matter superfields has been proposed and proved in \cite{Shifman:2014cya,Shifman:2015doa} in the case of using the higher covariant derivative regularization \cite{Slavnov:1971aw,Slavnov:1972sq,Krivoshchekov:1978xg,West:1985jx}. This equation is valid for the $D$-function and the anomalous dimension defined in terms of the bare coupling constant and has the form

\begin{equation}\label{ShifmanFormula}
D(\alpha_{s0})=\frac{3}{2}N\sum\limits_{\alpha=1}^{N_{f}} q_{\alpha}^2\Big(1-\gamma(\alpha_{s0})\Big),
\end{equation}

\noindent
where $N_{f}$ flavours of the matter superfields with electric charges $q_{\alpha}$ lie in the fundamental representation of $SU(N)$. The proof was based on the method used in \cite{Stepanyantz:2011jy} for the derivation of the NSVZ $\beta$-function  \cite{Novikov:1983uc,Jones:1983ip,Novikov:1985rd,Shifman:1986zi} defined in terms of the bare coupling constant in the ${\cal N}=1$ supersymmetric quantum electrodynamics regularized by higher covariant derivatives by direct summation of Feynman supergraphs. The key idea of this method is based on the following observation made in \cite{Soloshenko:2003nc,Smilga:2004zr}: For supersymmetric gauge theories regularized by higher derivatives, quantum contributions to the two-point Green function of the gauge superfield factorize into integrals of total and even double total derivatives in the momentum space in the limit of the vanishing external momentum. This allows performing explicit integration with respect to one loop momentum in a given contribution to the $\beta$-function in the $n$-loop approximation and relating it to a contribution to the anomalous dimension in the $(n-1)$-loop approximation. The idea was fully realized in two different ways in Refs. \cite{Stepanyantz:2011jy,Stepanyantz:2014ima}, where it was proved to all loops that the $\beta$-function is indeed given by integrals of double total derivatives with respect to a loop momentum. Due to this property, it is related to the anomalous dimension of the matter superfields through the NSVZ relation if both renormalization group (RG) functions are defined in terms of the bare coupling constant.\footnote{A similar factorization into double total derivatives takes place for integrals defining the renormalization of the photino mass in softly broken ${\cal N}=1$ SQED \cite{Nartsev:2016nym}.} These results were verified by explicit three-loop calculations in Ref. \cite{Kazantsev:2014yna}. Note that with dimensional reduction the factorization into double total derivatives does not take place \cite{Aleshin:2015qqc}, and the RG functions defined in terms of the bare coupling constant do not satisfy the NSVZ relation \cite{Aleshin:2016rrr}.

The formula (\ref{ShifmanFormula}) itself is an analogue of the NSVZ $\beta$-function for the ${\cal N}=1$ SYM theory with matter,

\begin{equation}\label{NSVZ}
\beta(\alpha)=-\frac{\alpha^2(3C_{2}-T(R)+C(R)_{i}{}^{j}\gamma(\alpha)_{j}{}^{i}/r)}{2\pi(1-C_{2}\alpha/(2\pi))},
\end{equation}

\noindent
where $\mbox{tr}(T^{A}T^{B})=T(R)\delta^{AB}$, $C(R)_i{}^j=(T^{A}T^{A})_i{}^j$, $C_{2}\delta^{CD}=f^{ABC}f^{ABD}$, $r=\delta^{AA}$.\footnote{In Eq. (\ref{NSVZ}) we do not yet specify the definitions of the RG functions and the argument of the NSVZ relation.}  The relation (\ref{ShifmanFormula}) even looks very similar to the exact NSVZ $\beta$-function for the ${\cal N}=1$ supersymmetric electrodynamics \cite{Vainshtein:1986ja,Shifman:1985fi}, but it is necessary to remember that the anomalous dimension on the right hand side is calculated in the non-Abelian theory.

Eq. (\ref{NSVZ}) has been derived by various methods, including instanton calculus \cite{Novikov:1983uc,Novikov:1985rd,Shifman:1999mv}, analysis of the anomaly supermultiplet \cite{Jones:1983ip,Shifman:1986zi,ArkaniHamed:1997mj}, and the non-renormalization theorem for the topological term \cite{Kraus:2002nu}. Explicit calculations in the ordinary perturbation theory, i.e. involving direct evaluation of contributions of multiloop Feynman diagrams, performed with dimensional reduction \cite{Siegel:1979wq,Siegel:1980qs} accompanied by the $\overline{\mbox{DR}}$ subtraction scheme, yielded the $\beta$-function in the three- \cite{Avdeev:1981ew,Jack:1996vg,Jack:1996cn,Harlander:2009mn} and even four-loop approximations \cite{Harlander:2006xq}. The result coincided with Eq. (\ref{NSVZ}) only in the two-loop approximation. In higher loops Eq. (\ref{NSVZ}) is satisfied only after a specially tuned finite renormalization of the coupling constant, which should be done in each order of perturbation theory \cite{Jack:1996vg,Jack:1996cn,Jack:1998uj}. Note that the very fact of its existence is quite non-trivial, as was noticed in \cite{Jack:1996vg}, because the NSVZ relation imposes certain scheme-independent restrictions on the divergences \cite{Kataev:2013csa,Kataev:2014gxa}. The general equations describing the scheme dependence of the NSVZ relation have been derived in \cite{Kutasov:2004xu,Kataev:2014gxa}.

At present, in the case of using dimensional reduction there is no general prescription leading to the subtraction scheme in which the NSVZ relation would hold to all orders (the NSVZ subtraction scheme). Moreover, dimensional reduction is mathematically inconsistent \cite{Siegel:1980qs}, and an attempt to remove the inconsistencies causes supersymmetry breaking by quantum corrections in higher loops \cite{Avdeev:1981vf,Avdeev:1982xy}. However, at present, it is not clear in which order this occurs. By contrast, the regularization by higher covariant derivatives appeared to be a remarkable computational tool for perturbative calculations in supersymmetric gauge theories, see e.g. \cite{Kazantsev:2014yna,Pimenov:2009hv,Stepanyantz:2011bz,Aleshin:2016yvj,Shakhmanov:2017soc}. It was first introduced in Refs. \cite{Slavnov:1971aw,Slavnov:1972sq}, and afterwards generalized to supersymmetric theories in \cite{Krivoshchekov:1978xg,West:1985jx}. It is mathematically consistent and preserves both gauge symmetry and supersymmetry. The use of the regularization by higher derivatives for multiloop calculations allowed constructing the NSVZ subtraction scheme in ${\cal N}=1$ SQED \cite{Kataev:2013eta}. To formulate the corresponding renormalization prescription, it is necessary to fix a value $x_{0}$  of $x = \ln(\Lambda/\mu)$, where $\Lambda$ is the dimensionful parameter of the regularized theory and $\mu$ is a renormalization scale, and require that the renormalization constants satisfy the condition

\begin{equation}\label{ConditionSQED}
Z(\alpha, x_{0})=1;\qquad Z_{3}(\alpha,x_{0})=1.
\end{equation}

\noindent
Then one can show that in this subtraction scheme the $\beta$-function and the anomalous dimension defined in terms of the renormalized coupling constant coincide with the corresponding RG functions defined in terms of the bare coupling constant in all orders of perturbation theory \cite{Kataev:2013eta}. Given that for the latter RG functions the NSVZ relation holds, its validity directly follows for the former ones. The non-Abelian analogue of the boundary conditions (\ref{ConditionSQED}) has been written in \cite{Stepanyantz:2016gtk}. It was verified by an explicit three-loop calculation for the terms quartic in the Yukawa couplings in \cite{Shakhmanov:2017soc}. This calculation is non-trivial, because the considered part of the three-loop $\beta$-function is scheme-dependent. The result exactly confirms the prescription for the NSVZ scheme given in \cite{Stepanyantz:2016gtk}. The subtraction scheme in which the photino mass renormalization in softly broken ${\cal N}=1$ SQED satisfies the NSVZ-like relation \cite{Hisano:1997ua,Jack:1997pa,Avdeev:1997vx} can also be constructed in a similar way \cite{Nartsev:2016mvn}. However, a subtraction scheme in which the exact expression for the Adler $D$-function is valid for the RG functions defined in terms of the renormalized coupling constant has not yet been found. Certainly, it would be interesting to construct such a scheme and to demonstrate how it works by an explicit calculation. Note that a non-trivial check can be obtained only starting from the three-loop approximation. The three-loop calculation is also a useful verification of Eq. (\ref{ShifmanFormula}), because its proof in Ref. \cite{Shifman:2015doa}, although exhaustive, is rather technical and complicated.

This paper is devoted to the calculation of the three-loop Adler $D$-function and the two-loop anomalous dimension for ${\cal N}=1$ SQCD regularized by higher covariant derivatives and to the verification of the relation (\ref{ShifmanFormula}) in the corresponding approximation. By an explicit calculation we prove the validity of Eq. (\ref{ShifmanFormula}) for the RG functions defined in terms of the bare coupling constant at three loops. Next, we construct the subtraction scheme in which the Adler $D$-function and the anomalous dimension defined in terms of the renormalized coupling constant satisfy Eq. (\ref{ShifmanFormula}) to all orders and illustrate it with a three-loop calculation. Our result differs from the one for ${\cal N}=1$ SQED obtained in \cite{Kataev:2013eta} only in that we have to impose the condition similar to (\ref{ConditionSQED}) on the renormalization constant for the strong coupling constant as well.

Let us comment on the version of the higher covariant derivative regularization we use in this work. Introduction of higher covariant derivatives into the action significantly modifies the vertices, which makes it hard to use this kind of regularization in practical calculations. But this circumstance is mitigated by the structure of quantum corrections in supersymmetric gauge theories regularized by higher derivatives, which we mentioned earlier. Moreover, calculations can be simplified even more, if one uses a version of regularization by higher derivatives breaking the BRST symmetry (a two-loop calculation with such a regularization was done in \cite{Pimenov:2009hv}). As a result, the Slavnov-Taylor identities have to be restored by the introduction of additional counterterms into the action (the corresponding renormalization procedure was constructed for both non-supersymmetric and supersymmetric Yang-Mills theories in \cite{Slavnov:2001pu,Slavnov:2002ir} and \cite{Slavnov:2002kg,Slavnov:2003cx}, respectively). On the other hand, it can sometimes be useful to sacrifice the convenience of calculations in favour of the BRST invariance. For example, a recent computation in \cite{Aleshin:2016yvj} with a BRST-preserving version of the higher derivative regularization in the Yang-Mills theory revealed the non-renormalization of the triple vertices with two ghost legs and one leg of the quantum gauge superfield at one loop, which later transformed into a non-renormalization theorem proved in all loops in \cite{Stepanyantz:2016gtk}, with the proof to a great extent relying on the Slavnov-Taylor identities. In this work we regularize the theory by higher covariant derivatives without breaking the BRST invariance, which considerably complicates the vertices but, nevertheless, leads to calculable expressions.

The paper is organized as follows: in Sect. \ref{SectionSQCD} we describe ${\cal N}=1$ SQCD (with the matter superfields in an arbitrary representation of the gauge group) and specify the version of the higher covariant derivative regularization which we will use throughout the calculations. In Sect. \ref{SectionThreeLoopD} we provide the three-loop expression for the Adler $D$-function written in terms of loop integrals. In Sect. \ref{SectionTwoLoopGamma} the result for the anomalous dimension in the two-loop approximation is presented as a sum of integrals and the validity of Eq. (\ref{ShifmanFormula}) is verified in the considered approximation. In Sect. \ref{SectionNSVZ} we construct the renormalization prescription fixing the subtraction scheme in which the relation (\ref{ShifmanFormula}) is valid to all loops for the RG functions defined in terms of the renormalized coupling constant. This prescription is verified by an explicit three-loop calculation in Sect. \ref{SectionThreeLoopNSVZ} after obtaining the final expressions for $\gamma(\alpha_{s0})$ and $D(\alpha_{s0})$ by evaluating the integrals for the simplest choice of the higher derivative regulator function. Feynman amplitudes for two- and three-loop diagrams contributing to the Adler $D$-function can be found in Appendix A, for one- and two-loop diagrams contributing to the anomalous dimension, in Appendix B. Appendix C is devoted to the evaluation of the integrals arising in the three-loop calculation.

\section{${\cal N}=1$ SQCD: the action, regularization, and renormalization}
\hspace*{\parindent}\label{SectionSQCD}

We will consider the ${\cal N}=1$ supersymmetric non-Abelian gauge theory with an arbitrary simple gauge group $G$ interacting with an external Abelian gauge field. The chiral matter superfields $\phi_\alpha$ and $\widetilde\phi_\alpha$ for each value of $\alpha =1,\ldots, N_f$ are assumed to lie in the representation $R+\bar R$. The bare action of the theory in the massless limit can be written as

\begin{eqnarray}\label{Theory}
&& S=\frac{1}{2g_{0}^2}\mbox{tr Re}\int d^4x d^2\theta W^{a} W_{a} + \frac{1}{4e_{0}^{2}}\mbox{Re}\int d^4x d^2\theta \bm{W}^{a} \bm{W}_{a}\nonumber\\
&&\qquad\qquad\qquad\qquad + \sum\limits_{\alpha=1}^{N_{f}} \frac{1}{4}\int d^4x d^4\theta \Big(\phi_{\alpha}^+ e^{2V+2q_{\alpha}\bm{V}}{\phi}_{\alpha} +{\widetilde{\phi}}_{\alpha}^+ e^{-2V^{t}-2q_{\alpha}\bm{V}}{\widetilde{\phi}}_{\alpha}\Big).\qquad
\end{eqnarray}

\noindent
This theory is invariant under the gauge transformations of the $G\times U(1)$ group. $V$ is the non-Abelian gauge superfield and $g_0$ is the corresponding bare coupling constant. Inside $W_a = \bar D^2(e^{-2V} D_a e^{2V})/8$ in the first term of Eq. (\ref{Theory}) $V=g_0 V^A t^A$, where $t^A$ are generators of the fundamental representation of the group $G$ normalized by the condition $\mbox{tr}(t^A t^B) = \delta^{AB}/2$. In the terms containing the matter superfields $V = g_0 V^A T^A$, where $T^A$ are generators of the representation $R$. The Abelian external classical gauge superfield is denoted by $\bm{V}$ and $e_0$ is the bare coupling constant corresponding to the $U(1)$ group. Each of the chiral matter superfields $\phi_\alpha$ belongs to the representation $R$ of the group $G$ and has the charge $q_\alpha$ with respect to $U(1)$. The superfields $\widetilde\phi_\alpha$ belong to the representation $\bar R$ and have the $U(1)$ charge $-q_\alpha$.

The regularization is implemented by adding to the action the term containing higher covariant derivatives,\footnote{It is not necessary to introduce a higher derivative term for the Abelian gauge superfield, because it is treated as a classical external superfield and can be present only on external legs.}

\begin{equation}\label{Regularization}
S_{\Lambda}=\frac{1}{2g_{0}^2}\mbox{tr Re}\int d^4x\, d^2\theta\, W^{a} \left[R\left(-\frac{{\bar{\nabla}}^2{\nabla}^2}{16{\Lambda}^2}\right)-1\right]W_{a},
\end{equation}

\noindent
where $\Lambda$ is the dimensionful parameter of the regularized theory playing the role of the ultraviolet cut-off. The function $R(y)$, such that $R(0)=1$, is the regulator which should rapidly grow at infinity, and we introduced the covariant derivatives in the chiral representation

\begin{equation}
\bar\nabla_{\dot{a}}=\bar D_{\dot{a}}\qquad\nabla_{a}=e^{-2V}D_{a}e^{2V}.
\end{equation}

\noindent
Note that regularizing terms of this type preserve BRST invariance \cite{Becchi:1974md,Tyutin:1975qk,Piguet:1981fb} of the total action after the gauge-fixing procedure.

We perform gauge fixing by adding the term

\begin{equation}\label{Gauge-Fixing}
S_{gf}=-\frac{1}{16\xi_0 g_0^2}\mbox{tr}\int d^4x\, d^4\theta\, \bar D^2 V R\big(\partial^2/\Lambda^2\big) D^2 V,
\end{equation}

\noindent
where $\xi_0$ is the bare gauge parameter. Due to the Slavnov--Taylor identities quantum corrections to the two-point Green function of the gauge superfield $V$ are transversal, so that the gauge fixing term is not renormalized. This implies that the renormalization of the gauge parameter $\xi$ is related to the renormalization of the coupling constant $g$ and of the superfield $V$. According to \cite{Aleshin:2016yvj}, the one-loop running of the gauge parameter is described by the equation

\begin{equation}
\frac{1}{\xi_0 g_0^2} = \frac{1}{\xi g^2} + \frac{C_2(1-\xi)}{12\xi \pi^2} \Big(\ln\frac{\Lambda}{\mu} + a_1\Big) + O(g^2),
\end{equation}

\noindent
where $\xi$ and $g$ are the renormalized gauge parameter and the renormalized coupling constant, respectively. Choosing, for simplicity, the Feynman gauge $\xi=1$, in the approximation considered in this paper we can make in the gauge fixing term the replacement $\xi_0 g_0^2 \to g^2$. Then the propagator of the gauge superfields $V^A$ is given by the expression

\begin{equation}\label{Propagator}
2i\left(\frac{1}{\partial^2 R}-\frac{g^2-g_0^2}{16 g_0^2}\Big(D^2 \bar D^2 + \bar D^2 D^2\Big)\frac{1}{\partial^4 R}\right)\delta^8(x_1-x_2) \delta^{AB}.
\end{equation}

\noindent
The action for the Faddeev-Popov ghosts has the form

\begin{equation}
S_{FP}= \frac{1}{g_0^2}\int d^4x\, d^4\theta\, (\bar c + \bar c^+)\Big[\Big(\frac{V}{1-e^{2V}}\Big)_{Adj}c^+ + \Big(\frac{V}{1-e^{-2V}}\Big)_{Adj}c\Big]
\end{equation}

\noindent
with the subscript $Adj$ implying that the ghosts lie in the adjoint representation of the non-Abelian group.

After adding the term $S_\Lambda$ to the action $S$ only the one-loop divergences remain \cite{Faddeev:1980be}. They should be regularized by the help of the Pauli--Villars method \cite{Slavnov:1977zf}. The one-loop divergences coming from the matter loop can be regularized by introducing a single set of anticommuting Pauli--Villars superfields $\Phi_\alpha$ and $\widetilde\Phi_\alpha$ with the action

\begin{equation}
S_{PV}=  \sum\limits_{\alpha=1}^{N_{f}}\Bigg[\frac{1}{4}\int d^4x\, d^4\theta\, \Big(\Phi_{\alpha}^{+} e^{2V+2q_{\alpha}\bm{V}}{\Phi}_{\alpha} + \widetilde{\Phi}_{\alpha}^{+} e^{-2V^{t}-2q_{\alpha}\bm{V}}{\widetilde{\Phi}}_{\alpha}\Big)
+\Big(\frac{1}{2}\int d^4x\, d^2\theta\, M \widetilde\Phi_{\alpha}^{t}{\Phi}_\alpha + \mbox{c.c.}\Big)\Bigg],
\end{equation}

\noindent
where $M=a\Lambda$ with $a$ being a dimensionless parameter independent of the coupling constant. Following Ref. \cite{Aleshin:2016yvj}, to cancel the divergences coming from the loops of the gauge superfield $V$ and of the Faddeev-Popov ghosts in the one-loop approximation, we introduce an additional set of (commuting) Pauli-Villars superfields with the action

\begin{eqnarray}
&& S_{\varphi}=\frac{1}{2}\mbox{tr}\int d^4x\, d^4\theta\, \varphi_{1}^{+} \Big[e^{2V}R\Big(-\frac{\bar\nabla^2{\nabla}^2}{16{\Lambda}^2}\Big)\Big]_{Adj}\varphi_{1}+\frac{1}{2}\mbox{tr}\int d^4x\, d^4\theta\, \varphi_{2}^{+}\big[e^{2V}\big]_{Adj}\varphi_{2}\qquad\nonumber\\
&& +\frac{1}{2}\mbox{tr}\int d^4x\, d^4\theta\, \varphi_{3}^{+} \left[e^{2V}\right]_{Adj}\varphi_{3}+\frac{1}{2}\Big(\mbox{tr}\int d^4x\, d^2\theta\, M_{\varphi}(\varphi_{1}^2+\varphi_{2}^2+\varphi_{3}^2)+\mbox{c.c.}\Big),
\end{eqnarray}

\noindent
where $M_\varphi = a_\varphi \Lambda$ and $a_\varphi$ is a parameter analogous to $a$ which does not depend on the coupling constant. Due to the gauge invariance of the action under the $U(1)$ gauge transformations, quantum corrections to the two-point Green function of  $\bm{V}$ are transversal,

\begin{equation}\label{DeltaGamma}
\Delta\Gamma^{(2)}_{\bm{V}} = -\frac{1}{16\pi}\int \frac{d^4p}{{(2\pi)}^4}\, d^4\theta\, \bm{V}(-p,\theta){\partial}^2{\Pi}_{1/2}\bm{V}(p,\theta)\,\Big(d^{-1}\big(\alpha_{0},\alpha_{s0}, \Lambda/p\big)-{\alpha}_{0}^{-1} \Big).
\end{equation}

\noindent
Here $\alpha_0=e_0^2/4\pi$ is the bare coupling constant corresponding to the $U(1)$ group, $\alpha_{s0}=g_0^2/4\pi$ is the bare coupling constant corresponding to the group $G$, and

\begin{equation}
\Pi_{1/2} = -\frac{D^a \bar D^2 D_a}{8 \partial^2}
\end{equation}

\noindent
is the transversal projection operator. Similarly, quantum corrections to the two-point Green function of the non-Abelian quantum gauge superfield are also transversal due to the BRST symmetry. This fact can be proved in the standard way by using the Slavnov--Taylor identities \cite{Taylor:1971ff,Slavnov:1972fg}.

Quantum contributions to the two-point Green function of the chiral matter superfields enter the effective action as

\begin{eqnarray}
&& \Gamma^{(2)}_{\phi}=\frac{1}{4}\sum\limits_{\alpha=1}^{N_{f}}\int \frac{d^4p}{(2\pi)^4}\, d^4\theta\, \Big(\phi_{\alpha}^{*i}(-p,\theta)G_{i}{}^{j}\big(\alpha_{s0},\Lambda/p\big)\phi_{\alpha j}(p,\theta)\nonumber\\
&&\qquad\qquad\qquad\qquad\qquad\qquad\qquad\qquad +\widetilde{\phi}_{\alpha j}^*(-p,\theta)G^{j}{}_{i}\big(\alpha_{s0},\Lambda/p\big)\widetilde{\phi}_\alpha^{i}(p,\theta)\Big).\qquad
\end{eqnarray}

The renormalized coupling constants $\alpha(\alpha_{0},\alpha_{s0}, \Lambda/\mu)$, $\alpha_{s}(\alpha_{s0},\Lambda/\mu)$ and the renormalization constants for the chiral matter superfields $Z_{i}{}^{j}(\alpha_{s},\Lambda/\mu)$, such that

\begin{equation}
\phi_{\alpha i}=(\sqrt{Z})_{i}{}^{j}(\phi_{R})_{\alpha j};\qquad \widetilde{\phi}_{\alpha }^{i}=(\sqrt{Z})^{i}{}_{j}(\widetilde \phi_{R})_\alpha^{j},
\end{equation}

\noindent
are defined in the standard way.\footnote{The renormalization constants for the superfields $\phi$ and $\widetilde\phi$ coincide, because the substitution $\phi \leftrightarrow \widetilde\phi^*$ is equivalent to the replacement $V \to - V$, $\bm{V} \to - \bm{V}$, and $D\leftrightarrow \bar D$.} Instead of the functions $\alpha(\alpha_{0},\alpha_{s0}, \Lambda/\mu)$ and $\alpha_{s}(\alpha_{s0},\Lambda/\mu)$ one can equivalently use the renormalization constants

\begin{equation}
Z_\alpha \equiv \alpha/\alpha_0;\qquad\mbox{and}\qquad Z_{\alpha_s} \equiv \alpha_s/\alpha_{s0},
\end{equation}

\noindent
respectively.

In the one-loop approximation the coupling constant $\alpha_{s}$ is related to the corresponding bare constant by the equation

\begin{equation}\label{One-LoopRunning}
\frac{1}{\alpha_{s0}}-\frac{1}{\alpha_{s}}=\frac{1}{2\pi}\Big[3C_{2}\Big(\ln\frac{\Lambda}{\mu}+b_{11}\Big) - 2 N_{f} T(R) \Big(\ln\frac{\Lambda}{\mu}+b_{12}\Big)\Big] + O(\alpha_{s}),
\end{equation}

\noindent
where $b_{11}$ and $b_{12}$ are constants whose particular values depend on the subtraction scheme.

Following \cite{Shifman:2014cya,Shifman:2015doa}, the Adler $D$-function in terms of the bare coupling constant is defined as

\begin{equation}\label{AdlerBareDefinition}
D(\alpha_{s0})=-\left.\frac{3\pi}{2}\frac{d}{d\ln\Lambda }\alpha_{0}^{-1}\left(\alpha,\alpha_{s},\Lambda/\mu\right)\right|_{\alpha,\alpha_{s}=\mbox{\scriptsize const}}.
\end{equation}

\noindent
This function can be expressed in terms of the two-point Green function of the Abelian gauge superfield $\bm{V}$,

\begin{equation}\label{HowToEvalD}
D(\alpha_{s0})= \frac{3\pi}{2} \frac{d}{d\ln\Lambda}\Big[d^{-1}\Big(\alpha_{0}(\alpha,\alpha_{s},\Lambda/\mu),\alpha_{s0}(\alpha_{s},\Lambda/\mu),\Lambda/p\Big)
-\alpha_{0}^{-1}(\alpha,\alpha_{s},\Lambda/\mu)\Big]\Bigg|_{\alpha,\alpha_{s}=\mbox{\scriptsize const};\,p=0}.
\end{equation}

\noindent
The equality follows from the fact that the derivative of the function $d^{-1}$ expressed in terms of the renormalized coupling constants with respect to $\ln\Lambda$, with $\alpha$ and $\alpha_{s}$  fixed, vanishes in the limit $p\rightarrow 0$. This limit is needed in order to get rid of the finite terms proportional to $(p/\Lambda)^k$, where $k$ is a positive integer.

Similarly, the anomalous dimension of each of the matter superfields $\phi_{\alpha}$ (which is the same for all $\alpha=1,\ldots,N_f$) defined in terms of the bare coupling constant can be also expressed in terms of the two-point Green function of the chiral matter superfields,

\begin{equation}\label{GammaBare}
\gamma_{i}{}^{j}(\alpha_{s0})=-\frac{d\ln Z_{i}{}^{j}}{d\ln\Lambda}\Big|_{\alpha_{s}=\mbox{\scriptsize const}} = \frac{d\ln G_{i}{}^{j}(\alpha_{s0}(\alpha_{s},\Lambda/\mu),\Lambda/q)}{d\ln\Lambda}\Big|_{\alpha_{s}=\mbox{\scriptsize const};\,q=0}.
\end{equation}

In this paper we argue that for the theory (\ref{Theory}) in the three-loop approximation the $D$-function and the anomalous dimension of the matter superfields are related by the equation

\begin{equation}\label{WhatWeWantToProve}
D(\alpha_{s0})=\frac{3}{2}\sum\limits_{\alpha=1}^{N_{f}}q_{\alpha}^{2}\Big(\mbox{dim}(R) - \mbox{tr}\,\gamma(\alpha_{s0})\Big),
\end{equation}

\noindent
where $\mbox{dim}(R)$ is the dimension of the representation $R$. This equation is the generalization of Eq. (\ref{ShifmanFormula}) to the considered case. Really, if the chiral matter superfields $\phi_\alpha$ belong to the fundamental representation of the $SU(N)$ group, then

\begin{equation}
\mbox{dim}(R) = N \qquad \mbox{and} \qquad \gamma_i{}^j(\alpha_{s0}) = \gamma(\alpha_{s0})\,\delta_i^j,
\end{equation}

\noindent
so that

\begin{equation}
\mbox{tr}(\gamma) = \gamma_i{}^i(\alpha_{s0}) = N \gamma(\alpha_{s0}).
\end{equation}

\noindent
Substituting these expressions into Eq. (\ref{WhatWeWantToProve}), we obtain Eq. (\ref{ShifmanFormula}) derived in \cite{Shifman:2014cya,Shifman:2015doa}.

\section{The three-loop Adler $D$-function}
\hspace*{\parindent}\label{SectionThreeLoopD}

\begin{center}
\begin{figure}[h]
\begin{tabular}{cccccccc}
\begin{minipage}{45pt}
\flushleft{(0)}\hfill\includegraphics[scale=0.066]{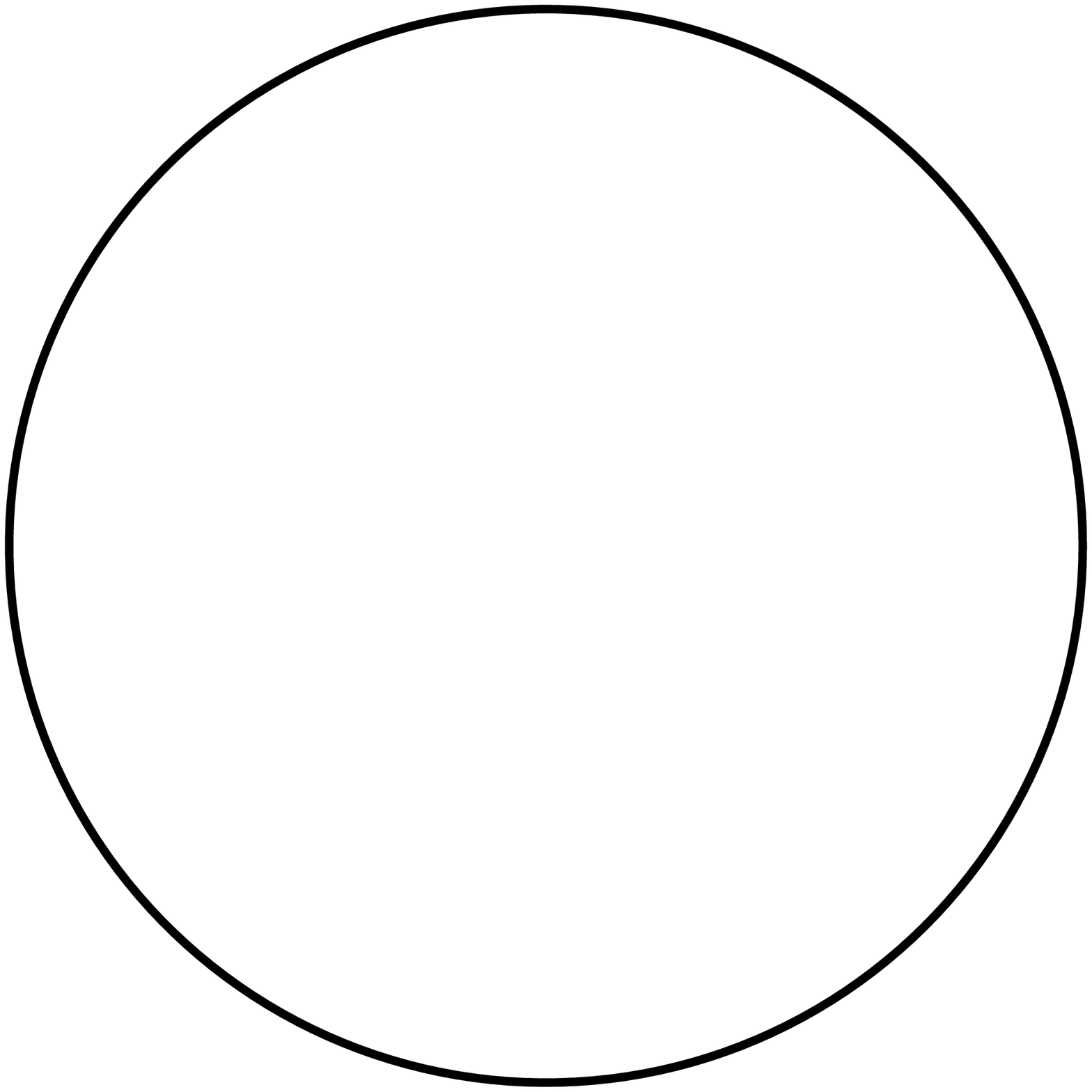}
\end{minipage}&
\begin{minipage}{45pt}
\flushleft{(1)}\hfill\includegraphics[scale=0.4]{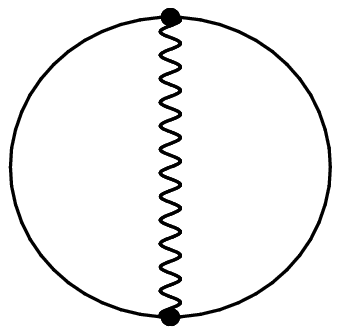}
\end{minipage}&
\begin{minipage}{45pt}
\flushleft{(2)}\hfill\includegraphics[scale=0.4]{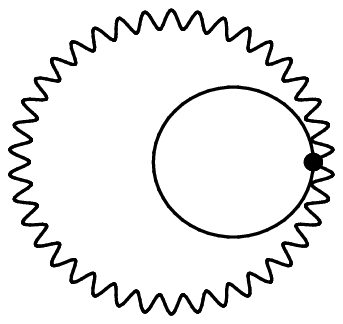}
\end{minipage}&
\begin{minipage}{45pt}
\flushleft{(3)}\hfill\includegraphics[scale=0.4]{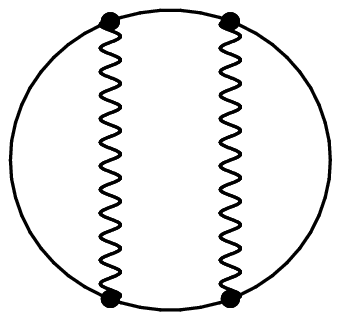}
\end{minipage}&
\begin{minipage}{45pt}
\flushleft{(4)}\hfill\includegraphics[scale=0.4]{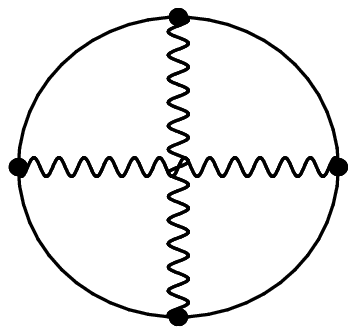}
\end{minipage}&
\begin{minipage}{45pt}
\flushleft{(5)}\hfill\includegraphics[scale=0.4]{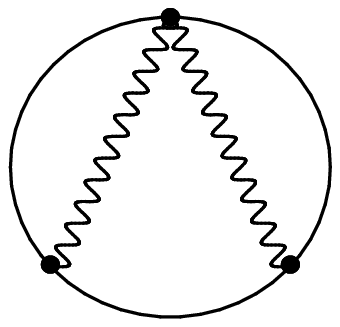}
\end{minipage}&
\begin{minipage}{45pt}
\flushleft{(6)}\hfill\includegraphics[scale=0.4]{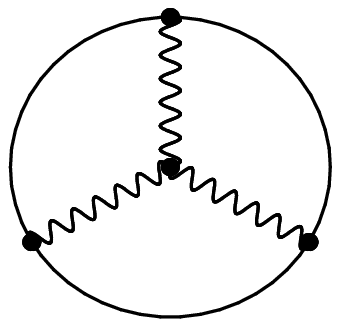}
\end{minipage}&
\begin{minipage}{45pt}
\flushleft{(7)}\hfill\includegraphics[scale=0.4]{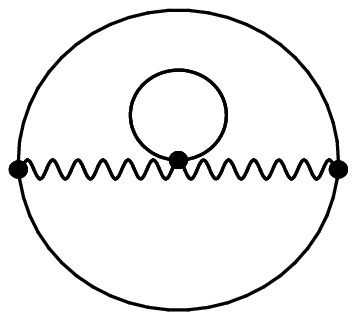}
\end{minipage}\\[1.5cm]
\begin{minipage}{45pt}
\flushleft{(8)}\hfill\includegraphics[scale=0.4]{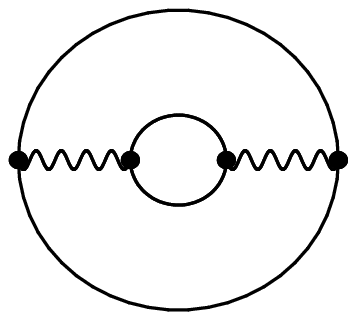}
\end{minipage}&
\begin{minipage}{45pt}
\flushleft{(9)}\hfill\includegraphics[scale=0.4]{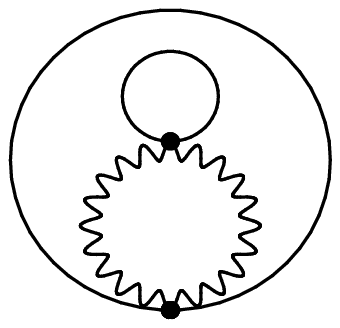}
\end{minipage}&
\begin{minipage}{45pt}
\flushleft{(10)}\hfill\includegraphics[scale=0.4]{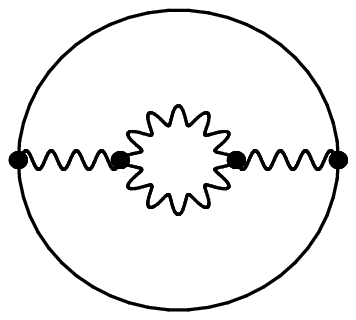}
\end{minipage}&
\begin{minipage}{45pt}
\flushleft{(11)}\hfill\includegraphics[scale=0.4]{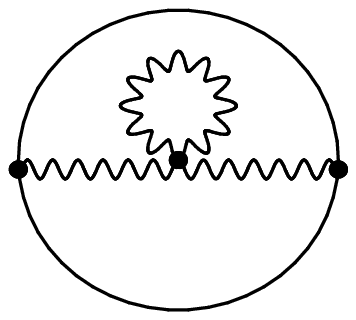}
\end{minipage}&
\begin{minipage}{45pt}
\flushleft{(12)}\hfill\includegraphics[scale=0.4]{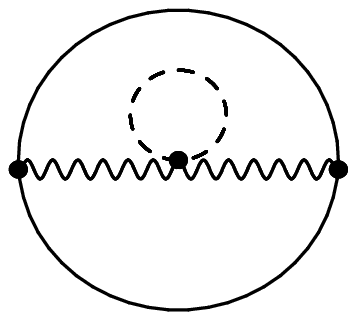}
\end{minipage}&
\begin{minipage}{45pt}
\flushleft{(13)}\hfill\includegraphics[scale=0.4]{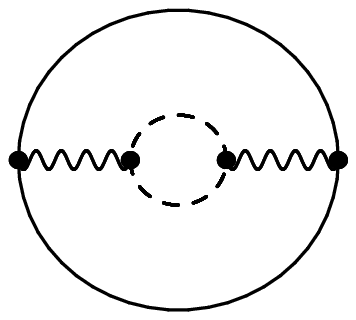}
\end{minipage}&
\begin{minipage}{45pt}
\flushleft{(14)}\hfill\includegraphics[scale=0.4]{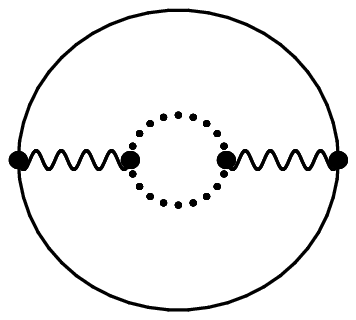}
\end{minipage}&
\begin{minipage}{45pt}
\flushleft{(15)}\hfill\includegraphics[scale=0.4]{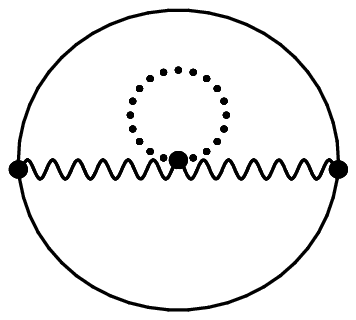}
\end{minipage}\\[1.5cm]
\begin{minipage}{45pt}
\flushleft{(16)}\hfill\includegraphics[scale=0.4]{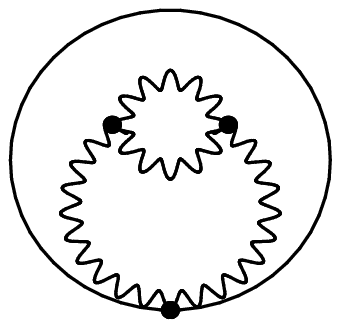}
\end{minipage}&
\begin{minipage}{45pt}
\flushleft{(17)}\hfill\includegraphics[scale=0.4]{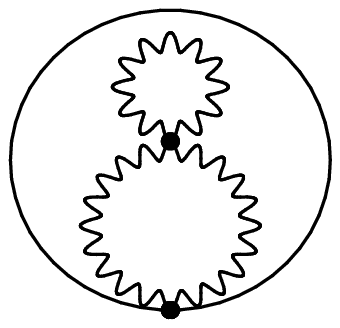}
\end{minipage}&
\begin{minipage}{45pt}
\flushleft{(18)}\hfill\includegraphics[scale=0.4]{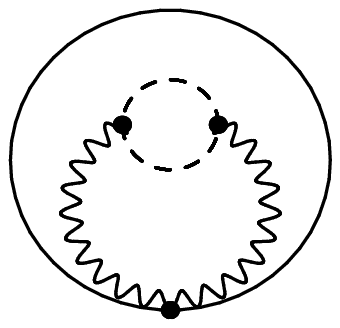}
\end{minipage}&
\begin{minipage}{45pt}
\flushleft{(19)}\hfill\includegraphics[scale=0.4]{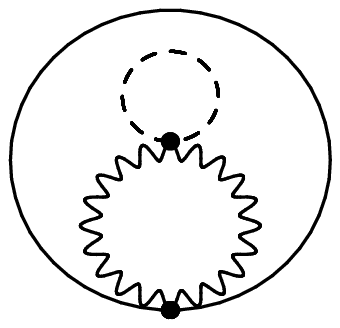}
\end{minipage}&
\begin{minipage}{45pt}
\flushleft{(20)}\hfill\includegraphics[scale=0.4]{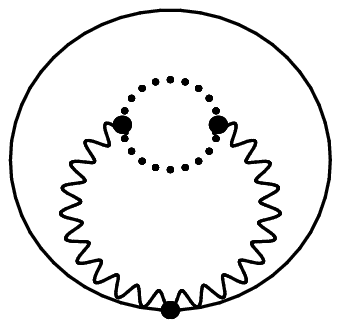}
\end{minipage}&
\begin{minipage}{45pt}
\flushleft{(21)}\hfill\includegraphics[scale=0.4]{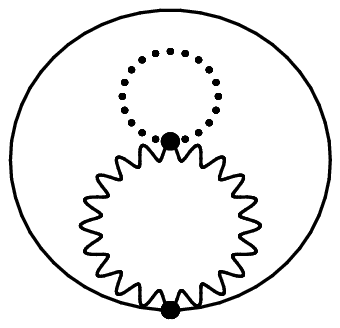}
\end{minipage}&
\begin{minipage}{45pt}
\flushleft{(22)}\hfill\includegraphics[scale=0.4]{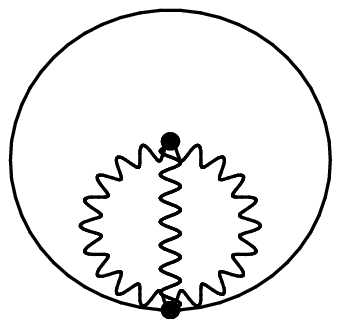}
\end{minipage}&
\begin{minipage}{45pt}
\flushleft{(23)}\hfill\includegraphics[scale=0.4]{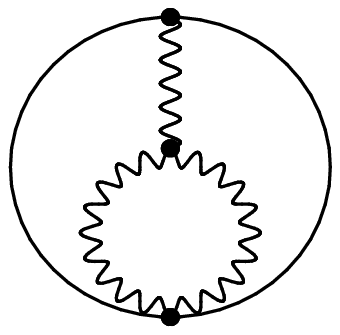}
\end{minipage}
\end{tabular}
\vspace*{3mm}
\caption{Two- and three-loop graphs contributing to the Adler $D$-function. Note that, for simplicity, some graphs vanishing in the considered (Feynman) gauge are not presented in this figure.}\label{Figure_Adler}
\end{figure}
\end{center}

According to Eq. (\ref{HowToEvalD}) the Adler function defined in terms of the bare coupling constant is related to the two-point Green function of the superfield $\bm{V}$. That is why the $D$-function is contributed to by superdiagrams with two external lines of the Abelian gauge superfield (which certainly do not contain its internal lines). There is a large number of such diagrams in the three-loop approximation. That is why we do not draw all of them in this paper. Instead of this, in Fig. \ref{Figure_Adler} we present only graphs without external lines. Attaching two external lines of the Abelian gauge superfield $\bm{V}$ in all possible ways to the loops of the matter superfields in these diagrams, we obtain the total three-loop contribution to the Adler $D$-function.

In Fig. \ref{Figure_Adler} the solid lines represent propagators of the matter superfields and the corresponding Pauli--Villars superfields, the wavy lines correspond to propagators of the non-Abelian gauge superfield $V$, and the dashed lines denote propagators of the Faddeev--Popov ghosts. The dotted lines represent propagators of the Pauli--Villars superfields $\varphi_i$. They are considered separately from the Pauli--Villars superfields $\Phi_\alpha$ and $\widetilde\Phi_\alpha$, because $\varphi_i$ do not interact with the external Abelian gauge superfield $\bm{V}$ unlike $\Phi_\alpha$ and $\widetilde\Phi_\alpha$.

To obtain the $D$-function in the considered approximation, it is necessary to attach two external lines of the superfield $\bm{V}$ to the solid lines in these graphs and
calculate the expression (\ref{HowToEvalD}) finding the function $d^{-1}-\alpha_0^{-1}$ by the help of Eq. (\ref{DeltaGamma}).

The results for contributions of the graphs presented in Fig. \ref{Figure_Adler} to Eq. (\ref{HowToEvalD}) are collected in Appendix \ref{AppendixGraphs}. All these expressions are given by integrals of double total derivatives. Their sum can be written as

\begin{eqnarray}\label{ThreeLoopD}
&& D(\alpha_{s0}) = \frac{3\pi}{2}\cdot 4\pi \sum\limits_{\alpha=1}^{N_f} q_\alpha^2 \int\frac{d^4q}{(2\pi)^4} \frac{\partial}{\partial q^\mu} \frac{\partial}{\partial q_\mu} \frac{d}{d\ln\Lambda} \Big[\,\mbox{dim}(R)\, I_0(q) + \mbox{tr}\, C(R)\, I_1(q) \qquad\nonumber\\
&& + \mbox{tr}\left(C(R)^2\right) I_2(q) + C_2\, \mbox{tr}\, C(R)\,\left( I_3(q) + I_4(q)\vphantom{)^2}\right) + N_f T(R)\, \mbox{tr}\, C(R)\, I_5(q)\Big],\qquad
\end{eqnarray}

\noindent
where $I_i(q)$ with $i=1,\ldots, 4$ are given by the differences of the massless contributions of the usual matter superfields $\phi$, $\widetilde\phi$ and the massive contributions of the Pauli--Villars superfields $\Phi$, $\widetilde\Phi$,

\begin{equation}
I_i(q) = I_i(q, m=0) - I_i(q,m=M)\qquad i=1,2,3,4.
\end{equation}

\noindent
The contribution $I_5(q)$ comes from the graphs (7), (8) and (9), which have two loops of the matter superfields, and has a different structure. The explicit expressions for the integrals $I_1(q,m),\ldots, I_4(q,m)$, $I_5(q)$, and the one-loop contribution $I_0(q)$ have the form

\begin{eqnarray}
&&\hspace*{-3mm} I_0(q) = \frac{1}{2q^2} \ln\Big(1+\frac{M^2}{q^2}\Big);\\
&&\hspace*{-3mm} I_1(q,m) = \int \frac{d^4k}{(2\pi)^4}\,\frac{g_0^2}{R_k k^2 \big(q^2+m^2\big)\big((q+k)^2+m^2\big)};\\
&&\hspace*{-3mm} I_2(q,m) =  \int \frac{d^4k}{(2\pi)^4}\,\frac{d^4l}{(2\pi)^4}\, \frac{g_0^4}{k^2 R_k l^2 R_l \big(q^2+m^2\big) \big((q+k)^2+m^2\big)\big((q+l)^2+m^2\big)}\nonumber\\
&&\hspace*{-3mm}\qquad\qquad\qquad\qquad\qquad\qquad\qquad\quad \times \Big(\frac{2(q^2-m^2)}{q^2+m^2} + \frac{(2q+k+l)^2+2m^2}{(q+k+l)^2 + m^2} - 4\Big);\qquad\\
&&\hspace*{-3mm} I_3(q,m) = \int \frac{d^4k}{(2\pi)^4}\,\frac{d^4l}{(2\pi)^4}\, \frac{g_0^4}{k^2 R_k l^2 R_l \big(q^2+m^2\big) \big((q+k)^2+m^2\big)\big((q+l)^2+m^2\big)}\nonumber\\
&&\hspace*{-3mm} \times \left(-\frac{(2q+k+l)^2+2m^2}{2\big((q+k+l)^2 + m^2\big)} + 1 - \frac{2\big((q+k)^\mu (q+l)_\mu + m^2\big)}{(l-k)^2} - \frac{4}{(l-k)^2 R_{l-k}} \frac{R_l-R_k}{l^2-k^2}\right.\nonumber\\
&&\hspace*{-3mm}\left. \times  \Big( q^2 (q+k)^\mu l_\mu + l^2 (q+k)^\mu q_\mu + m^2 (q+k+l)^\mu l_\mu\Big)\vphantom{\frac{1}{2}}\right);\vphantom{\frac{1}{2}}\\
&&\hspace*{-3mm} I_4(q,m) = - \int \frac{d^4k}{(2\pi)^4}\, \frac{2 g_0^4 f(k/\Lambda)}{R_k^2 k^2 \big(q^2+m^2\big)\big((q+k)^2 + m^2\big)};\\
&&\hspace*{-3mm} I_5(q) = - \int \frac{d^4k}{(2\pi)^4}\,\frac{d^4l}{(2\pi)^4}\, \frac{2 g_0^4}{k^2 R_k^2}\Big(\frac{1}{q^2(q+k)^2} - \frac{1}{\big(q^2+M^2\big)\big((q+k)^2+M^2\big)} \Big)\nonumber\\
&&\hspace*{-3mm}\qquad\qquad\qquad\qquad\qquad\qquad\qquad\qquad \times \Big(\frac{1}{l^2(k+l)^2} - \frac{1}{\big(l^2+M^2\big)\big((k+l)^2+M^2\big)} \Big).
\end{eqnarray}

\noindent
The function $f(k/\Lambda)$ entering the expression for $I_4(q,m)$ is related to the one-loop contribution to the polarization operator of the quantum gauge superfield and has been calculated in Ref. \cite{Kazantsev:2017fdc}. It is given by Eq. (\ref{Polarization}) in Appendix \ref{AppendixGraphs}.

From Eq. (\ref{ThreeLoopD}) we see that in the considered approximation the Adler $D$-function is given by integrals of double total derivatives with respect to the momentum $q^\mu$ of the matter loop. This exactly agrees with the results of Refs. \cite{Shifman:2014cya,Shifman:2015doa}. Such a structure of loop integrals allows taking the integral over $d^4q$. Note that although being integrals of total derivatives, the contributions to the $D$-function do not vanish due to singularities of the integrands. When evaluating each of the integrals, one should surround singular points by spheres of infinitely small radii and transform the volume integral into a sum of surface integrals over these spheres and over the infinitely large sphere. The last one can be discarded if the integrand decreases at infinity rapidly enough,

\begin{eqnarray}\label{IntegrationFormula}
&& \int\frac{d^4q}{(2\pi)^4}\frac{\partial}{\partial q^{\mu}}\frac{\partial}{\partial q_{\mu}}\frac{F(q^2)}{q^2}=\lim_{R\to\infty}\oint\limits_{S_{R}}\frac{dS}{(2\pi)^4}\frac{q^{\mu}}{q}\frac{\partial}{\partial q^{\mu}}\frac{F(q^2)}{q^2}-\lim_{\varepsilon\to 0}\oint\limits_{S_{\varepsilon}}\frac{dS}{(2\pi)^4}\frac{q^\mu}{q}\frac{\partial}{\partial q^\mu}\frac{F(q^2)}{q^2}\nonumber\\
&& =\frac{1}{4\pi^2}\lim_{q\to\infty}(F'(q^2)q^2-F(q^2))-\frac{1}{4\pi^2}\lim_{q\to 0}(F'(q^2)q^2-F(q^2))=\frac{F(0)}{4\pi^2},
\end{eqnarray}

\noindent
where the function $F(q^2)$ is assumed to be non-singular and rapidly decreasing at infinity. Using this equation it is possible to take all integrals over $d^4q$ in Eq. (\ref{ThreeLoopD}). Then, after some transformations, the result for the Adler $D$-function can be written in the form

\begin{eqnarray}\label{FinalIntegral}
&& D(\alpha_{s0}) =\frac{3}{2}\sum\limits_{\alpha=1}^{N_{f}}q_{\alpha}^2\Bigg[\mbox{dim}(R) + \frac{d}{d\ln\Lambda}\Bigg(\mbox{ tr}\,C(R) \int
\frac{d^4k}{(2\pi)^4}\frac{2g_{0}^2}{k^4 R_{k}}\Big(1- 2 C_2 \frac{g_0^2 f(k/\Lambda)}{R_k} \Big)
\qquad \nonumber\\
&& - N_{f}T(R)\,\mbox{tr}\,C(R) \int\frac{d^4k}{(2\pi)^4}\frac{d^4l}{(2\pi)^4}\frac{4g_{0}^4}{k^4R_{k}^2}\Big(\frac{1}{l^2 (k+l)^2} - \frac{1}{(l^2+M^2)((k+l)^2+M^2)}\Big)
\nonumber\\
&& + \mbox{tr}\left(C(R)^2\right) \int\frac{d^4k}{(2\pi)^4}\frac{d^4l}{(2\pi)^4}\frac{2g_{0}^4}{k^2R_{k}l^2R_{l}}\Big(\frac{1}{l^2k^2}-\frac{2}{l^2(l+k)^2}\Big)\Bigg) \Bigg] + O(g_0^6).
\end{eqnarray}

\noindent
Note that in Eq. (\ref{FinalIntegral}) the differentiation with respect to $\ln\Lambda$ should precede the integration. It is necessary in order to get rid of the  infrared divergences and to obtain well-defined integrals. This differentiation should be done with the renormalized coupling constant $\alpha_{s}$  fixed, as is required by Eq. (\ref{HowToEvalD}).

\section{The two-loop anomalous dimension of the matter superfields}
\hspace*{\parindent}\label{SectionTwoLoopGamma}

\begin{figure}[h]
\begin{picture}(0,15)
\put(0.1,14.5){$(1)$}
\put(0.2,13){\includegraphics[scale=0.37]{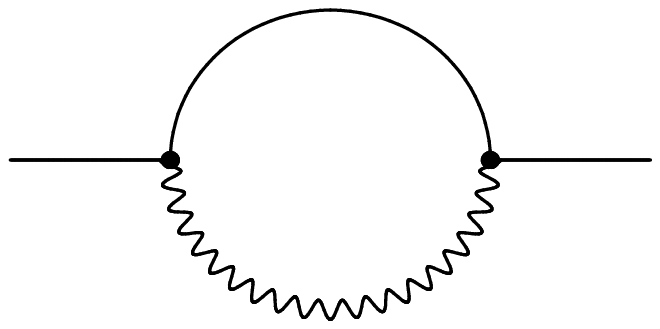}}
\put(3.4,14.5){$(2)$}
\put(3.5,13){\includegraphics[scale=0.37]{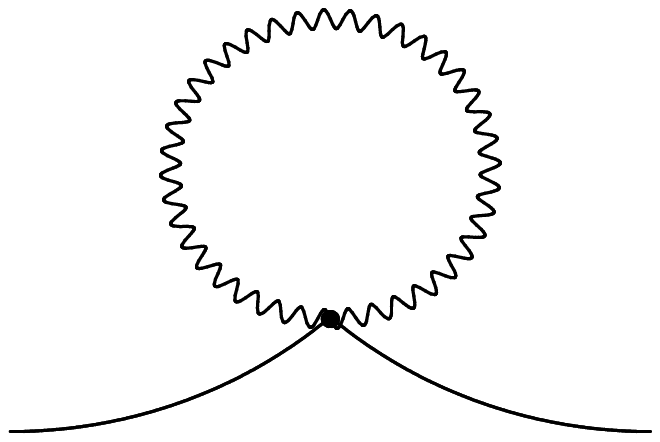}}
\put(6.7,14.5){$(3.1)$}
\put(6.8,13){\includegraphics[scale=0.37]{dga}}
\put(10.0,14.5){$(3.2)$}
\put(10.1,13){\includegraphics[scale=0.37]{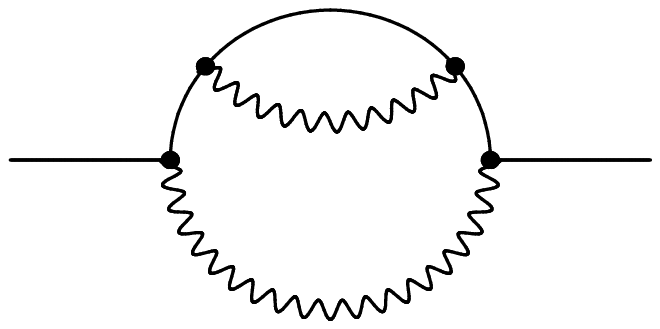}}
\put(13.3,14.5){$(4)$}
\put(13.4,13){\includegraphics[scale=0.37]{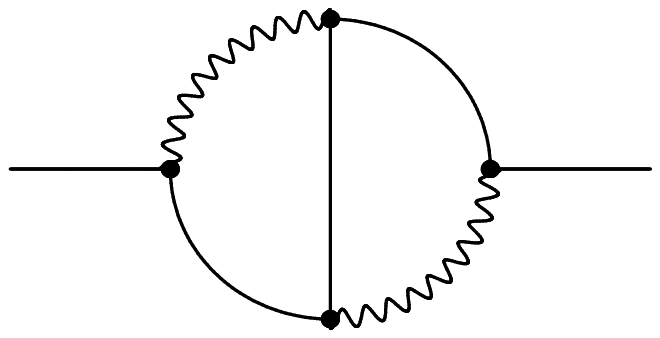}}

\put(0.1,12){$(5.1)$}
\put(0.2,10.5){\includegraphics[scale=0.37]{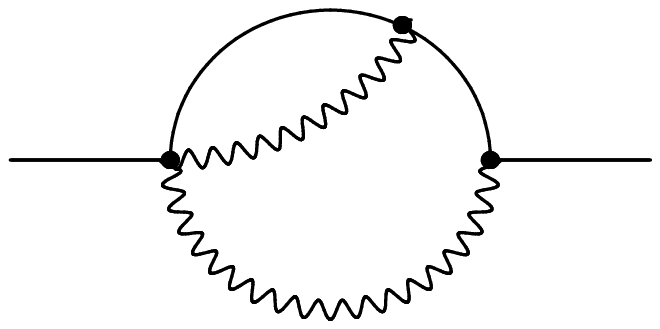}}
\put(3.8,12){$(5.2)$}
\put(3.9,10.5){\includegraphics[scale=0.37]{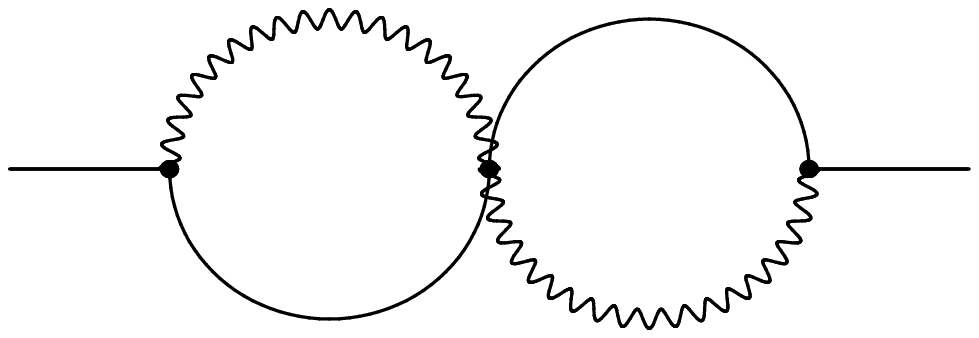}}
\put(8.9,12){$(6)$}
\put(9.0,10.9){\includegraphics[scale=0.5]{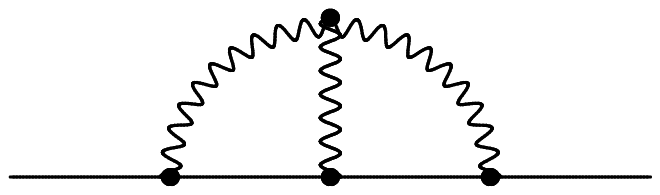}}
\put(13.3,12){$(7.1)$}
\put(13.4,10.0){\includegraphics[scale=0.37]{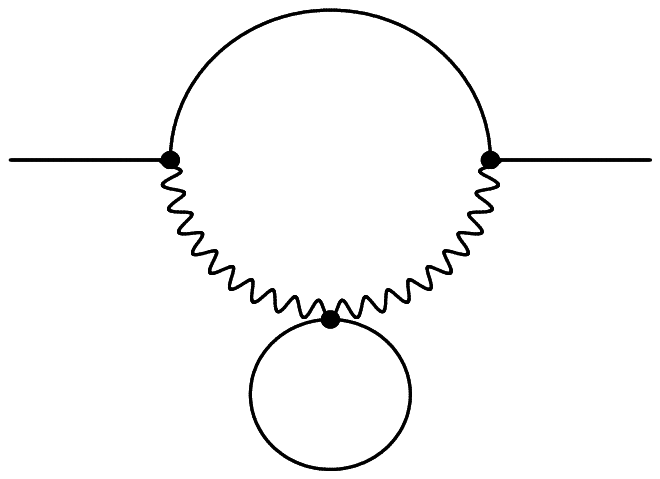}}

\put(0.1,9.5){$(7.2)$}
\put(0.2,7.9){\includegraphics[scale=0.37]{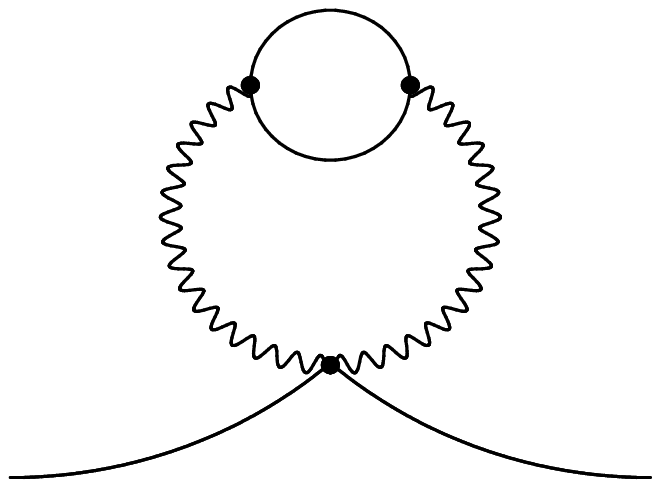}}
\put(3.4,9.5){$(8)$}
\put(3.5,8){\includegraphics[scale=0.37]{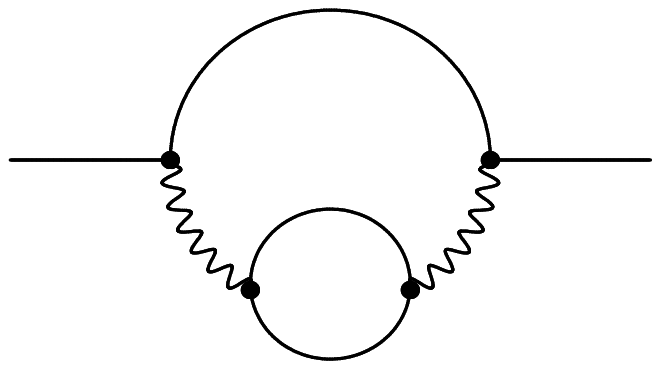}}
\put(6.7,9.5){$(9)$}
\put(6.8,7.8){\includegraphics[scale=0.37]{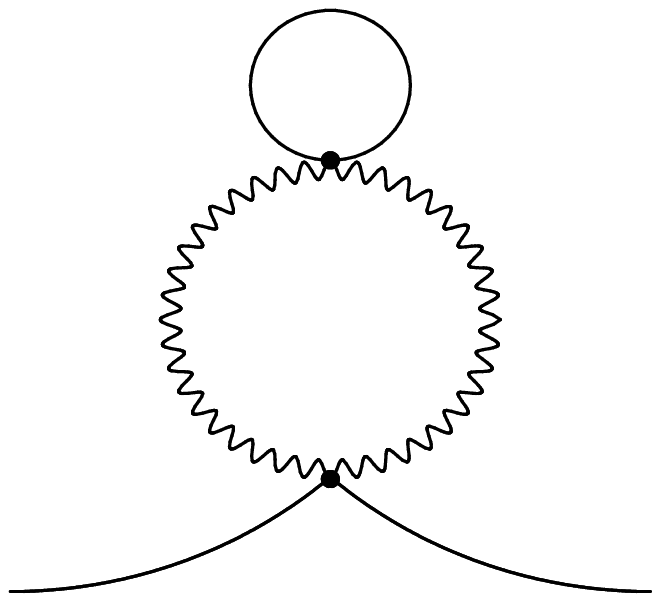}}
\put(10.0,9.5){$(10)$}
\put(10.1,8){\includegraphics[scale=0.37]{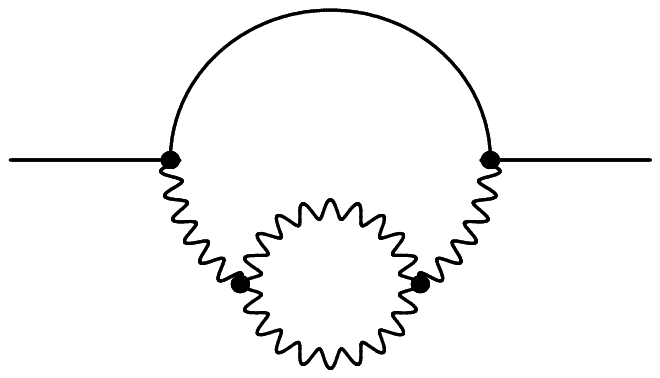}}
\put(13.3,9.5){$(11)$}
\put(13.4,7.6){\includegraphics[scale=0.37]{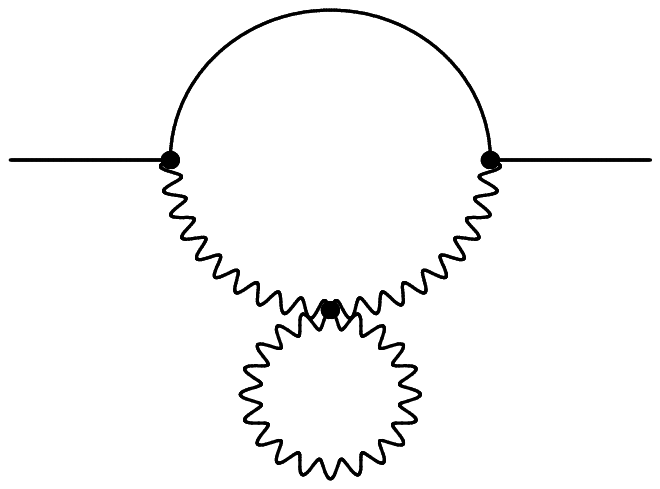}}

\put(0.1,7.0){$(12)$}
\put(0.2,5.1){\includegraphics[scale=0.37]{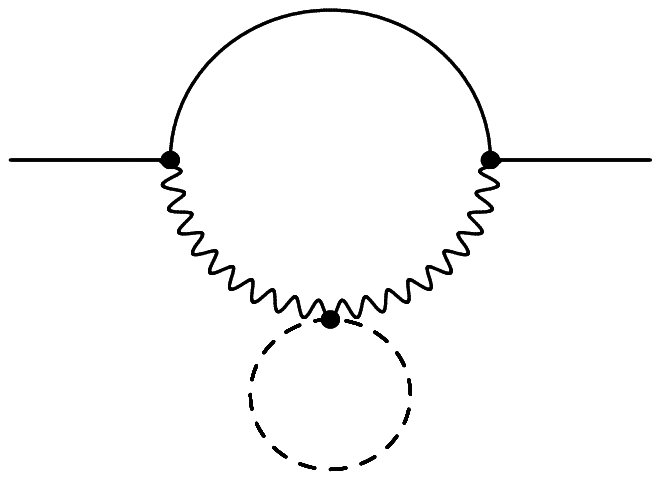}}
\put(3.4,7.0){$(13)$}
\put(3.5,5.5){\includegraphics[scale=0.37]{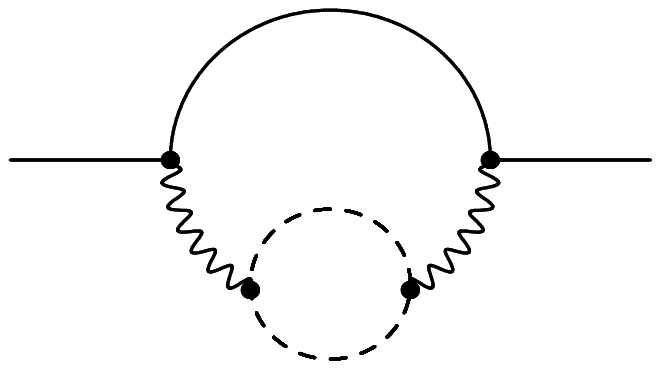}}
\put(6.7,7.0){$(14)$}
\put(6.8,5.5){\includegraphics[scale=0.37]{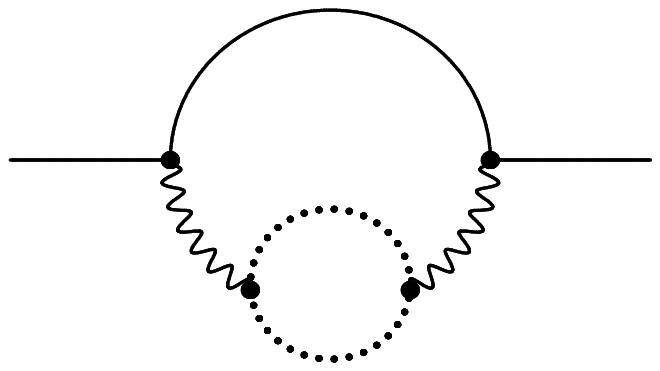}}
\put(10.0,7.0){$(15)$}
\put(10.1,5.1){\includegraphics[scale=0.37]{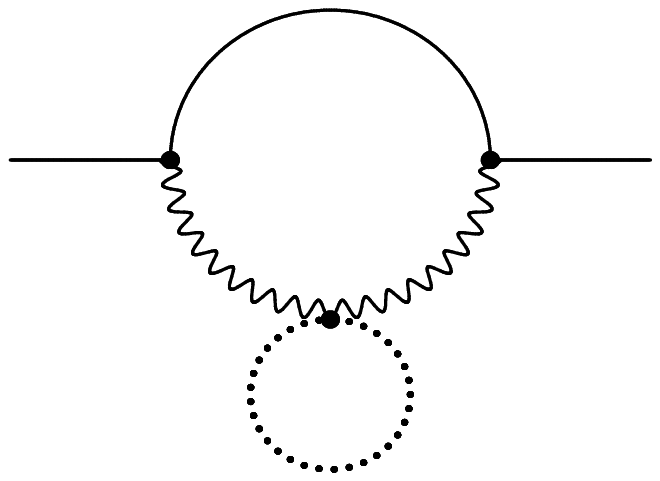}}
\put(13.3,7.0){$(16)$}
\put(13.4,5.2){\includegraphics[scale=0.37]{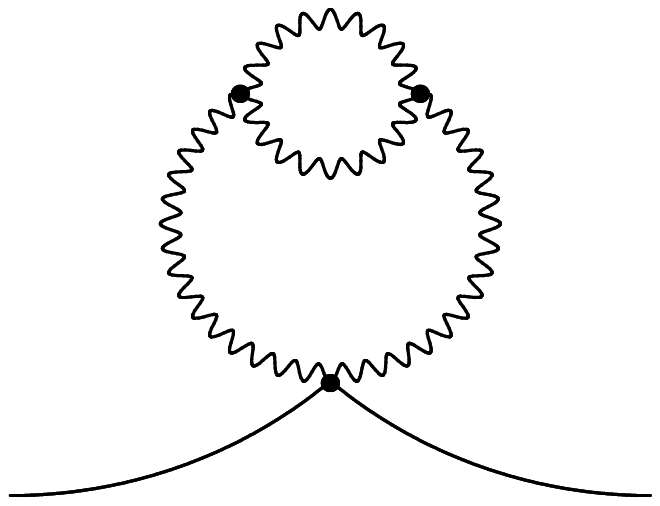}}

\put(0.1,4.5){$(17)$}
\put(0.2,2.5){\includegraphics[scale=0.37]{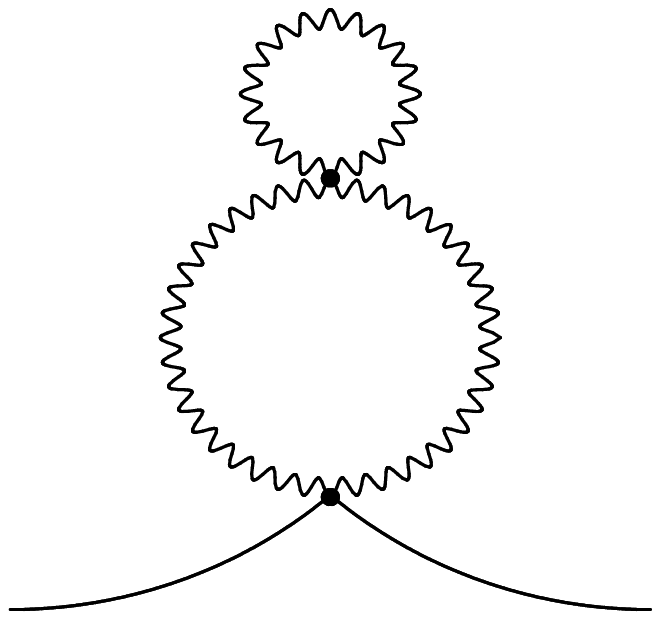}}
\put(3.4,4.5){$(18)$}
\put(3.5,2.5){\includegraphics[scale=0.37]{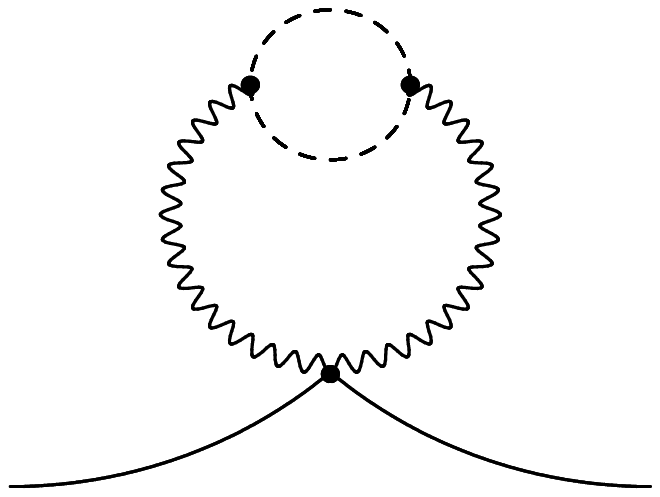}}
\put(6.7,4.5){$(19)$}
\put(6.8,2.5){\includegraphics[scale=0.37]{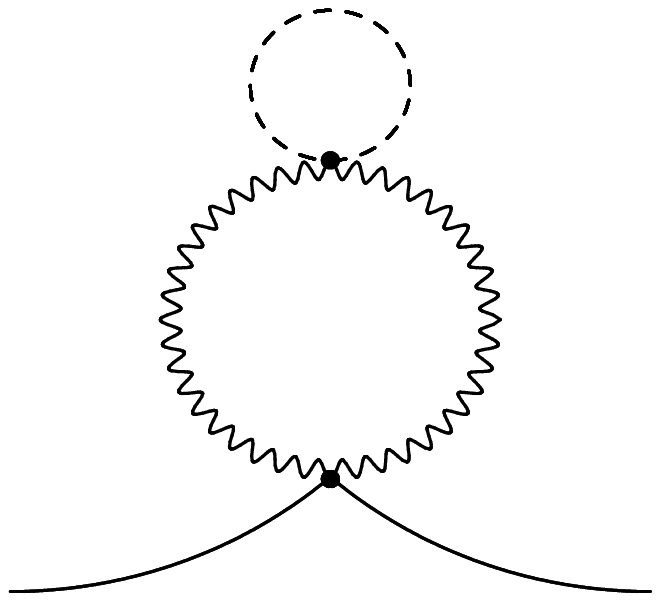}}
\put(10.0,4.5){$(20)$}
\put(10.1,2.5){\includegraphics[scale=0.37]{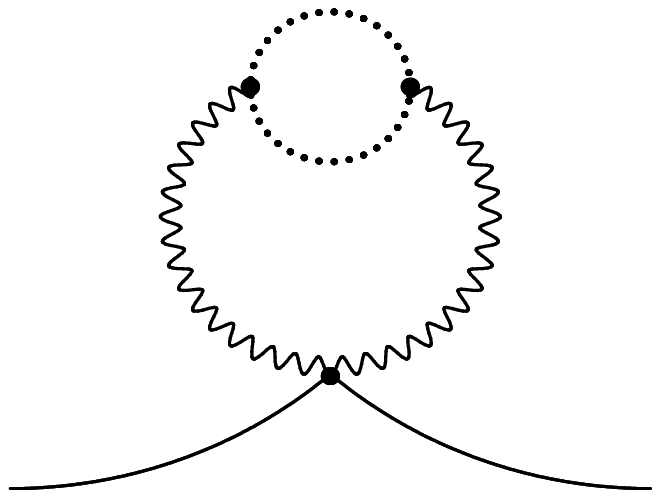}}
\put(13.3,4.5){$(21)$}
\put(13.4,2.5){\includegraphics[scale=0.37]{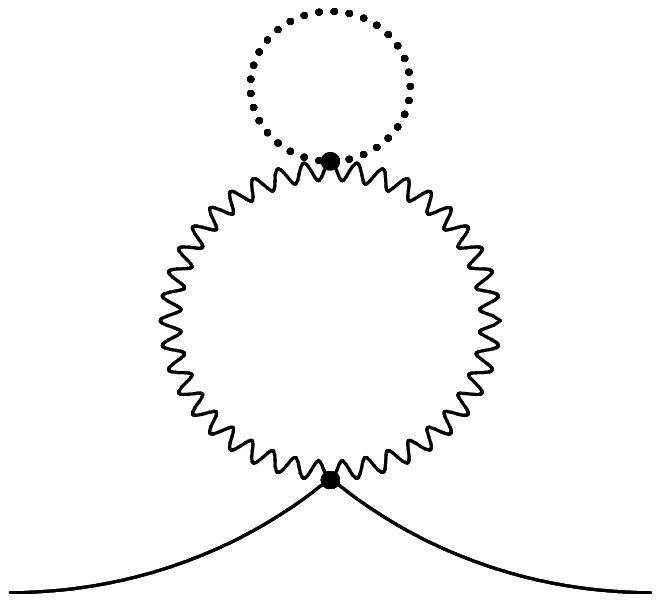}}

\put(4.8,1.8){$(22)$}
\put(4.9,0.1){\includegraphics[scale=0.37]{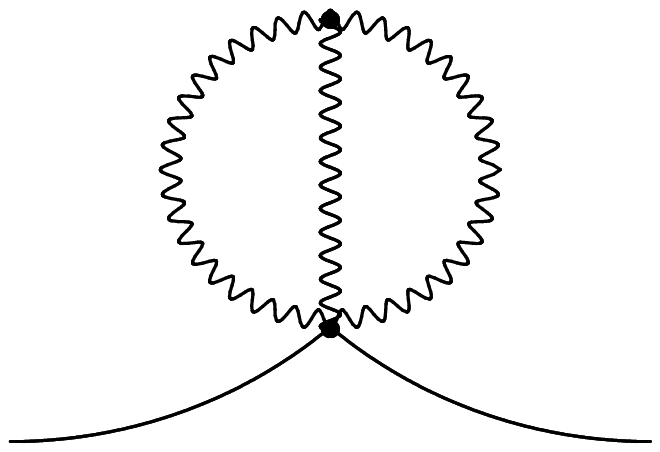}}
\put(9.0,1.8){$(23)$}
\put(9.1,0.1){\includegraphics[scale=0.5]{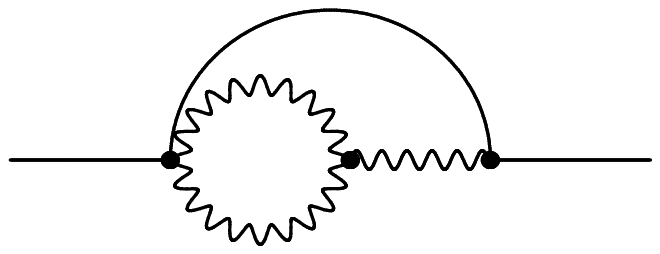}}
\end{picture}

\caption{Diagrams contributing to the two-loop anomalous dimension. Each graph in this figure can be obtained from the corresponding graph in Fig. \ref{Figure_Adler} by cutting a line of the matter superfields. For simplicity, some evidently vanishing diagrams are omitted exactly as in Fig. \ref{Figure_Adler}.}\label{Figure_Gamma}
\end{figure}

According to Eq. (\ref{WhatWeWantToProve}) the three-loop Adler function is related to the two-loop anomalous dimension of the matter superfields. Therefore, to verify this equation it is necessary to calculate the two-loop contribution to the two-point Green function of the matter superfields. The corresponding superdiagrams are presented in Fig. \ref{Figure_Gamma}. Note that they can be obtained from the graphs in Fig. \ref{Figure_Adler} by cutting the matter loops in as many ways as possible. As we are calculating their contribution to the effective action, only one particle irreducible diagrams resulting from this procedure should be kept. The expressions for quantum corrections received by $G(\alpha_{s0},\Lambda/q)$ from these diagrams are collected in Appendix \ref{AppendixMatter}. Using them, the anomalous dimension defined in terms of the bare coupling constant $\alpha_{s0}$ is calculated by the help of Eq. (\ref{GammaBare}). Namely, it is necessary to differentiate the sum of all contributions to $\ln G$ with respect to $\ln\Lambda$ and take the limit of the vanishing external momentum. This gives

\begin{eqnarray}\label{FinalIntegralGamma}
&&\hspace*{-5mm} \gamma_i{}^j(\alpha_{s0}) = - C(R)_i{}^j \frac{d}{d\ln\Lambda} \int\frac{d^4k}{(2\pi)^4}\frac{2g_{0}^2}{k^4R_{k}} + C_{2} C(R)_i{}^j \frac{d}{d\ln\Lambda}\int\frac{d^4k}{(2\pi)^4} \frac{4g_{0}^4}{k^4R_{k}^2}f(k/\Lambda)\nonumber\\
&&\hspace*{-5mm} + N_{f} T(R) C(R)_i{}^j \frac{d}{d\ln\Lambda} \int\frac{d^4k}{(2\pi)^4}\frac{d^4l}{(2\pi)^4}\frac{4g_{0}^4}{k^4R_{k}^2} \Big(\frac{1}{l^2 (l+k)^2} - \frac{1}{\big(l^2+M^2\big)\big((l+k)^2+M^2\big)}\Big) \nonumber\\
&&\hspace*{-5mm} - \left(C(R)^2\right)_i{}^j \frac{d}{d\ln\Lambda} \int\frac{d^4k}{(2\pi)^4}\frac{d^4l}{(2\pi)^4} \frac{2g_{0}^4}{k^2R_{k}l^2R_{l}}\Big(\frac{1}{l^2k^2} -\frac{2}{l^2(l+k)^2}\Big) + O\left(g_{0}^6\right),
\end{eqnarray}

\noindent
where the function $f(k/\Lambda)$ is the same as in (\ref{FinalIntegral}) and is given by Eq. (\ref{Polarization}).

Comparing Eqs. (\ref{FinalIntegral}) and (\ref{FinalIntegralGamma}) one can easily verify that the NSVZ-like relation (\ref{WhatWeWantToProve}) is really valid in the considered approximation for the RG functions defined in terms of the bare coupling constant $\alpha_{s0}$,

\begin{equation}\label{ThreeLoopRelation}
D(\alpha_{s0})=\frac{3}{2}\sum\limits_{\alpha=1}^{N_{f}} q_{\alpha}^2\Big(\mbox{dim}(R) - \mbox{tr}\, \gamma(\alpha_{s0})\Big) + O(\alpha_{s0}^3).
\end{equation}

\noindent
The equality takes place, because both sides of this equation are given by the same integrals, so it is valid independently of the subtraction scheme. This follows from the fact \cite{Kataev:2013eta} that the RG functions defined in terms of the bare coupling constant are scheme-independent for a fixed regularization. This can be confirmed by an explicit calculation in the considered approximation, for the simplest form of the regulator $R(y)$ (see Eq. (\ref{Regularization})). Let us choose it in the form

\begin{equation}\label{ExplicitRegulator}
R(y) = 1 + y^n,
\end{equation}

\noindent
where $n$ is a positive integer. Then it is possible to find explicit expressions for the RG functions entering Eq. (\ref{ThreeLoopRelation}) calculating the integrals in Eq. (\ref{FinalIntegralGamma}). Details of this calculation are described in Appendix \ref{AppendixIntegrals}, and the result has the form

\begin{eqnarray}\label{TwoLoopGammaBare}
&& \gamma(\alpha_{s0})_i{}^j = - \frac{\alpha_{s0}}{\pi} C(R)_i{}^j - \frac{3\alpha_{s0}^2}{2\pi^2} C_2 C(R)_i{}^j  \Big(\ln a_\varphi + 1\Big) + \frac{\alpha_{s0}^2}{\pi^2} N_{f} T(R) C(R)_i{}^j \Big(\ln a + 1\Big)\nonumber\\
&& + \frac{\alpha_{s0}^2}{2\pi^2} \left(C(R)^2\right)_i{}^j  + O(\alpha_{s0}^3),
\end{eqnarray}

\noindent
where

\begin{equation}
a \equiv \frac{M}{\Lambda};\qquad a_\varphi \equiv \frac{M_\varphi}{\Lambda}.
\end{equation}

\noindent
Therefore, from Eq. (\ref{ThreeLoopRelation}) we conclude that for the regulator (\ref{ExplicitRegulator}) the three-loop Adler $D$-function defined in terms of the bare coupling constant is given by the expression

\begin{eqnarray}\label{ThreeLoopDBare}
&& D(\alpha_{s0}) = \frac{3}{2} \sum\limits_{\alpha=1}^{N_{f}}q_{\alpha}^2 \Big[ \mbox{dim}(R) + \frac{\alpha_{s0}}{\pi} \mbox{tr}\,C(R) + \frac{3\alpha_{s0}^2}{2\pi^2}\, C_{2}\, \mbox{tr}\,C(R) \Big(\ln a_\varphi + 1\Big) \quad\nonumber\\
&& - \frac{\alpha_{s0}^2}{\pi^2}\, N_{f}T(R)\, \mbox{tr}\, C(R) \Big(\ln a  +1\Big) - \frac{\alpha_{s0}^2}{2\pi^2}\, \mbox{tr}\left( C(R)^2\right) \Big] + O(\alpha_{s0}^3).
\end{eqnarray}

\noindent
We see that finite constants defining the subtraction scheme (such as $b_{11}$ and $b_{12}$ in Eq. (\ref{One-LoopRunning})) do not enter Eqs. (\ref{TwoLoopGammaBare}) and (\ref{ThreeLoopDBare}). Consequently, these functions are scheme independent (for a fixed regularization) in agreement with the general statement proved in \cite{Kataev:2013eta}.

\section{The NSVZ-like subtraction scheme}
\hspace*{\parindent}\label{SectionNSVZ}

The RG functions entering the relation (\ref{WhatWeWantToProve}) are defined in terms of the bare coupling constant and, therefore, do not depend on the subtraction scheme for a fixed regularization \cite{Kataev:2013eta}.  On the other hand, RG functions are usually defined as functions of renormalized couplings \cite{Bogolyubov:1980nc} and depend on the renormalization prescription. This implies that the relation (\ref{WhatWeWantToProve}) is not valid for the Adler $D$-function and the anomalous dimension defined in terms of the renormalized coupling constant in an arbitrary subtraction scheme. We will show that the scheme in which (\ref{WhatWeWantToProve}) holds for the RG functions defined in terms of the renormalized coupling constant (the NSVZ-like scheme) can be constructed by imposing simple conditions on the renormalization constants. Namely, let us fix a value $x_{0}$ of $\ln(\Lambda/\mu)$ and impose the requirement that the renormalization constants satisfy the equations

\begin{equation}\label{Condition}
Z(\alpha_{s},x_{0})_i{}^j=\delta_i{}^j;\qquad Z_{\alpha}(\alpha,\alpha_{s},x_{0})=1;\qquad Z_{\alpha_{s}}(\alpha_{s},x_{0})=1.
\end{equation}

\noindent
In the particular case $x_0=0$ this is an analogue of minimal subtractions (see, e.g., \cite{Collins:1984xc}) in the sense that only terms proportional to $\big(\ln\Lambda/\mu\big)^n$, with $n\ge 1$ being a positive integer, survive in the renormalization constants. We will call this scheme HDMS (i.e. Higher Derivatives $+$ Minimal Subtractions).

First, we prove that such a scheme exists. For convenience, let us introduce a new variable $x=\ln(\Lambda/\mu)$. Then, we consider some renormalization prescription and find  the functions $\alpha_{0}(\alpha,\alpha_{s},x)$, $\alpha_{s0}(\alpha_{s},x)$ and $Z(\alpha_{s}, x)_i{}^j$ for it. In general, the conditions (\ref{Condition}) are not satisfied, so that

\begin{equation}\label{Boundary}
\alpha_{0}(\alpha,\alpha_{s}, x_{0})=a(\alpha,\alpha_{s});\qquad \alpha_{s0}(\alpha_{s},x_{0})=b(\alpha_{s});\qquad Z(\alpha_{s},x_{0})_i{}^j =g(\alpha_{s})_i{}^j.
\end{equation}

\noindent
Evidently, for a fixed finite value of $x_0$ the functions $a(\alpha,\alpha_{s})$, $b(\alpha_{s})$ and $g(\alpha_{s})_i{}^j$ are finite. Then we choose new renormalized coupling constants $\alpha'$, $\alpha_s'$  and new renormalization constants for the chiral matter superfields $Z'{}_i{}^j$ such that

\begin{equation}\label{FiniteRenormalization}
\alpha'(\alpha, \alpha_{s})=a(\alpha,\alpha_{s});\qquad
\alpha'_{s}(\alpha_{s})=b(\alpha_{s});\qquad
Z'(\alpha'_{s},x)_i{}^j =g^{-1}\big(\alpha_{s}(\alpha'_{s})\big)_i{}^k\, Z\big(\alpha_{s}(\alpha'_{s}),x\big)_k{}^j.
\end{equation}

\noindent
By construction, the renormalized Green functions expressed in terms of the renormalized coupling constants $\alpha$ and $\alpha_s$ are finite. Certainly, they remain finite after the finite renormalization (\ref{FiniteRenormalization}). Furthermore, from Eqs. (\ref{Boundary}) and (\ref{FiniteRenormalization}) we conclude that the renormalization constants $Z'_\alpha = \alpha'/\alpha_0$, $Z'_{\alpha_s} = \alpha'_{s}/\alpha_{s0}$ and $Z'(\alpha'_s,x)_i{}^j$ satisfy the conditions (\ref{Condition}). This implies that the scheme fixed by the conditions (\ref{Condition}) exists. Therefore, it only remains to show that in this scheme (\ref{WhatWeWantToProve}) is valid for the RG functions defined in terms of the renormalized coupling constant.

The Adler $D$-function and the anomalous dimension are defined in terms of the renormalized coupling constant by the equations

\begin{eqnarray}\label{DTilde}
&& \widetilde{D}(\alpha_{s})= -\frac{3\pi}{2}\frac{d}{d\ln\mu}\alpha^{-1}(\alpha_{0},\alpha_{s0},\Lambda/\mu) \Big|_{\alpha_{0},\alpha_{s0}=\mbox{\scriptsize const}};\\ \label{GammaTilde}
&& \widetilde{\gamma}(\alpha_{s})_i{}^j = \frac{d\ln Z(\alpha_{s},\Lambda/\mu)_i{}^j}{d\ln\mu}\Big|_{\alpha_{s0}=\mbox{\scriptsize const}}.
\end{eqnarray}

\noindent
From (\ref{GammaTilde}) we obtain

\begin{equation}\label{GammaTildeEquation}
\widetilde{\gamma}(\alpha_{s})_i{}^j = - \frac{d\ln Z(\alpha_{s},x)_i{}^j}{dx}\Big|_{\alpha_{s}=\mbox{\scriptsize const}} - \frac{\partial\ln Z(\alpha_{s},x)_i{}^j}{\partial\alpha_{s}}\, \frac{\partial \alpha_{s}(\alpha_{s0},x)}{\partial x}.
\end{equation}

\noindent
Let us set $x=x_0$ in this equation. Then, the second term in this expression vanishes, because according to the first equation in (\ref{Condition})

\begin{equation}\label{PartialZed}
\frac{\partial\ln Z(\alpha_{s},x_{0})_i{}^j}{\partial\alpha_{s}}=0.
\end{equation}

\noindent
Moreover, the third equation in (\ref{Condition}) gives $\alpha_s(\alpha_{s0},x_0)=\alpha_{s0}$. Therefore, from Eq. (\ref{GammaTildeEquation}) we obtain

\begin{equation}
\widetilde\gamma(\alpha_s)_i{}^j\Big|_{\alpha_s = \alpha_{s0}} = \gamma(\alpha_{s0})_i{}^j.
\end{equation}

\noindent
This implies that both definitions of the anomalous dimension give the same function if the boundary conditions (\ref{Condition}) are satisfied.

Similar statement can be proved for the $D$-function. For this purpose we rewrite the definition (\ref{DTilde}) in the equivalent form

\begin{equation}
\widetilde{D}(\alpha_{s})= \frac{3\pi}{2\alpha}\frac{d}{d\ln\mu}\ln Z_\alpha(\alpha,\alpha_{s},\Lambda/\mu) \Big|_{\alpha_{0},\alpha_{s0}=\mbox{\scriptsize const}}.
\end{equation}

\noindent
As earlier, the differentiation gives

\begin{eqnarray}
&& \widetilde{D}(\alpha_{s}) = - \frac{3\pi}{2\alpha} \left(\frac{d\ln Z_\alpha(\alpha,\alpha_{s},x)}{dx}\Big|_{\alpha,\alpha_{s}=\mbox{\scriptsize const}} + \frac{\partial\ln Z_\alpha(\alpha,\alpha_{s},x)}{\partial\alpha}\, \frac{\partial\alpha(\alpha_0,\alpha_{s0},x)}{\partial x}\right.\qquad\nonumber\\
&&\left. + \frac{\partial\ln Z_\alpha(\alpha,\alpha_{s},x)}{\partial\alpha_{s}}\, \frac{\partial\alpha_{s}(\alpha_{s0},x)}{\partial x}\right).
\end{eqnarray}

\noindent
Setting $x=x_0$, we see that the last two terms vanish due to the boundary conditions (\ref{Condition}), while $\alpha_s(\alpha_{s0},x_0) = \alpha_{s0}$ and $\alpha(\alpha_0,\alpha_{s0},x_0) = \alpha_0$. Consequently, we obtain

\begin{equation}
\widetilde{D}(\alpha_{s})\Big|_{\alpha_s = \alpha_{s0}} = - \frac{3\pi}{2\alpha_0} \frac{d\ln Z_\alpha(\alpha,\alpha_{s},x)}{dx}\Big|_{\alpha,\alpha_{s}=\mbox{\scriptsize const}} = D(\alpha_{s0}).
\end{equation}

\noindent
Therefore, under the boundary conditions (\ref{Condition}) the RG functions defined in terms of the renormalized coupling constants coincide with the ones defined in terms of the bare coupling constants. Taking into account that the latter functions satisfy the NSVZ-like relation in the case of using the higher covariant derivative regularization, we conclude that the former ones also satisfy it in the subtraction scheme specified by the prescription (\ref{Condition}). This implies that Eq. (\ref{Condition}) produces the NSVZ-like scheme in all orders of perturbation theory with the higher covariant derivative regularization.

\section{The NSVZ-like scheme in the three-loop approximation}
\hspace*{\parindent}\label{SectionThreeLoopNSVZ}

Let us verify the general statements considered in the previous section by the explicit three-loop calculation. We will start with Eq. (\ref{TwoLoopGammaBare}) and integrate the RG equation (\ref{GammaBare}). Solving this equation for $\ln Z$ and taking into account Eq. (\ref{One-LoopRunning}), which describes the one-loop running of the coupling constant $\alpha_s$, we obtain

\begin{eqnarray}\label{Zed}
&& \ln Z_i{}^j = \frac{\alpha_{s}}{\pi}\, C(R)_i{}^j \Big(\ln\frac{\Lambda}{\mu} + g_{1}\Big) + \frac{3\alpha_{s}^2}{4\pi^2}\, C_2 C(R)_i{}^j \Big[-\ln^2\frac{\Lambda}{\mu}+2\ln\frac{\Lambda}{\mu}\Big(\ln a_\varphi + 1 - b_{11}\Big) \nonumber\\
&& + g_{21}\Big] - \frac{\alpha_{s}^2}{2\pi^2}\, N_{f} T(R) C(R)_i{}^j \Big[-\ln^2\frac{\Lambda}{\mu}+2\ln\frac{\Lambda}{\mu}\Big(\ln a + 1 - b_{12}\Big) + g_{22}\Big]
- \frac{\alpha_{s}^2}{2\pi^2} \left(C(R)^2\right)_i{}^j\quad\nonumber\\
&& \times \Big[\ln\frac{\Lambda}{\mu}+g_{23}\Big] + O(\alpha_{s}^3),
\end{eqnarray}

\noindent
where $g_{1}$, $g_{21}$, $g_{22}$, and $g_{23}$ are arbitrary finite constants.\footnote{Three different constants $g_{21}$, $g_{22}$, and $g_{23}$ in the two-loop approximation appear due to three different group theory factors.} Existence of these arbitrary constants follows from arbitrariness of choosing a subtraction scheme. Fixing values of these constants and the other similar constants, one fixes the subtraction scheme.

The finite constants do not enter the RG functions defined in terms of the bare coupling constant, but they are present in the scheme dependent expressions for the RG functions defined in terms of the renormalized coupling constant. To calculate the anomalous dimension defined as a function of the renormalized coupling, one needs to rewrite the equation (\ref{Zed}) in terms of $\alpha_{s0}$, differentiate it with respect to $\ln\mu$, and express the result in terms of $\alpha_{s}$. This gives

\begin{eqnarray}\label{TwoLoopGammaRenormalized}
&& \widetilde{\gamma}(\alpha_{s})_i{}^j =\frac{d\ln Z_i{}^j}{d\ln\mu}\Big|_{\alpha_{s0}=\mbox{\scriptsize const}}= - \frac{\alpha_{s}}{\pi}\, C(R)_i{}^j +\frac{\alpha_s^2}{\pi^2}\Big[ - \frac{3}{2}\, C_2\, C(R)_i{}^j \Big(\ln a_\varphi + 1 + g_{1} - b_{11}\Big)\quad \nonumber\\
&& + N_{f} T(R)\, C(R)_i{}^j  \Big( \ln a + 1 + g_{1} - b_{12}\Big) + \frac{1}{2}\left(C(R)^2\right)_i{}^j\Big] + O(\alpha_{s}^3).
\end{eqnarray}

\noindent
The presence of the finite constants $g_1$, $b_{11}$ and $b_{12}$ confirms the scheme-dependence of this RG function.

Eq. (\ref{TwoLoopGammaRenormalized}) can be compared with the result obtained with dimensional reduction in the $\overline{\mbox{DR}}$-scheme, see \cite{Jack:1996vg} and references therein. In our notation it is written as

\begin{equation}
\widetilde\gamma_{\overline{\mbox{\tiny DR}}}(\alpha_s)_i{}^j = - \frac{\alpha_{s}}{\pi}\, C(R)_i{}^j +\frac{\alpha_s^2}{\pi^2}\Big[- \frac{3}{4}\, C_2\, C(R)_i{}^j + \frac{1}{2}\, N_{f} T(R)\, C(R)_i{}^j  + \frac{1}{2} \left(C(R)^2\right)_i{}^j\Big] + O(\alpha_{s}^3).
\end{equation}

\noindent
We see that the scheme-independent terms proportional to $C(R)^2$ coincide, while the other terms coincide if we choose the values of finite constants satisfying the equations

\begin{equation}
\ln a_\varphi + 1 + g_{1} - b_{11} = \frac{1}{2};\qquad \ln a + 1 + g_{1} - b_{12} = \frac{1}{2}.
\end{equation}

\noindent
This implies that our result agrees with the result of \cite{Jack:1996vg}, because the difference in the regularization can always be compensated by a proper choice of the subtraction scheme.

Integrating the RG equation (\ref{AdlerBareDefinition}) (keeping $\alpha_{s}$ fixed) one can relate the renormalized coupling constant $\alpha$ (corresponding to the $U(1)$ group) to the bare coupling constant $\alpha_0$,

\begin{eqnarray}\label{EquationForAlpha}
&&\hspace*{-5mm} \alpha_{0}^{-1} - \alpha^{-1} = -\frac{1}{\pi}\sum\limits_{\alpha=1}^{N_{f}}q_{\alpha}^2 \Bigg(\mbox{dim}(R)\Big(\ln\frac{\Lambda}{\mu}+d_{1}\Big) + \frac{\alpha_{s}}{\pi}\, \mbox{tr}\,C(R) \Big(\ln\frac{\Lambda}{\mu}+d_{2}\Big) - \frac{3\alpha_{s}^2}{2\pi^2}\, C_{2}\, \mbox{tr}\,C(R) \nonumber\\
&&\hspace*{-5mm} \times \Big[\frac{1}{2}\ln^2\frac{\Lambda}{\mu}-\ln\frac{\Lambda}{\mu}\Big(\ln a_\varphi + 1 - b_{11}\Big) + d_{31}\Big] + \frac{\alpha_{s}^2}{\pi^2}\, N_{f} T(R)\, \mbox{tr}\, C(R) \Big[\frac{1}{2}\ln^2\frac{\Lambda}{\mu}  -\ln\frac{\Lambda}{\mu}\Big(\ln a + 1 \nonumber\\
&&\hspace*{-5mm} - b_{12}\Big) + d_{32}\Big] - \frac{\alpha_{s}^2}{2\pi^2}\, \mbox{tr}\left( C(R)^2\right) \Big(\ln\frac{\Lambda}{\mu} + d_{33}\Big) \Bigg) + O(\alpha_{s}^3).
\end{eqnarray}

\noindent
As we have already mentioned, the arbitrary constants $d_{1}$, $d_{2}$, $d_{31}$, $d_{32}$, and $d_{33}$, which appeared here as constants of integration, together with $g_{1}$, $g_{21}$, $g_{22}$, $g_{23}$, $b_{11}$, and $b_{12}$, fix the subtraction scheme in the considered approximation.

From Eq. (\ref{EquationForAlpha}) we can obtain the renormalization constant $Z_{\alpha} = \alpha/\alpha_0$.

Expressing the right hand side of (\ref{EquationForAlpha}) in terms of $\alpha_{s0}$, differentiating both sides with respect to $\ln\mu$, and then returning back to $\alpha_{s}$ we obtain the Adler $D$-function defined in terms of the renormalized coupling constant,

\begin{eqnarray}\label{ThreeLoopDRenormalized}
&&\hspace*{-5mm} \widetilde{D}(\alpha_{s}) = -\frac{3\pi}{2}\frac{d\alpha^{-1}}{d\ln\mu}\Big|_{\alpha_{0},\alpha_{s0}=\mbox{\scriptsize const}} =\frac{3}{2}\sum\limits_{\alpha=1}^{N_{f}}q_{\alpha}^2\Bigg(\mbox{dim}(R) + \frac{\alpha_{s}}{\pi}\, \mbox{tr}\,C(R) +\frac{\alpha_s^2}{\pi^2}\Big[ - \frac{1}{2}\, \mbox{tr} \left(C(R)^2\right) \nonumber\\
&&\hspace*{-5mm} + \frac{3}{2}\, C_2\, \mbox{tr}\,C(R)\,  \Big(\ln a_\varphi + 1 + d_2 - b_{11}\Big) - N_{f} T(R)\,\mbox{tr}\,C(R)\, \Big( \ln a + 1 + d_2 - b_{12} \Big)\Big] + O(\alpha_{s}^3)\Bigg).\nonumber\\
\end{eqnarray}

\noindent
Due to the presence of the finite constants this expression is scheme-dependent.\footnote{Note that the constants $d_{31}$, $d_{32}$, and $d_{33}$ are not essential in the considered approximation.} However, from Eqs. (\ref{TwoLoopGammaRenormalized}) and (\ref{ThreeLoopDRenormalized}) we see that the terms proportional to $\left(C(R)^2\right)_i{}^j$ in the two-loop anomalous dimension and the terms proportional to $\mbox{tr}\left(C(R)^2\right)$ in the $D$-function are scheme independent. The scheme-independence of such terms in various RG functions was discussed in Refs. \cite{Broadhurst:1999he,Kataev:2010tm,Kataev:2013vua,Kataev:2013csa} in the Abelian case and in Ref. \cite{Kataev:2014gxa} in the non-Abelian case.

Now, let us fix the values of the finite constants by imposing the boundary conditions (\ref{Condition}). First, we fix a value $x_0$ and consider the equation $Z_{\alpha_s}(\alpha_s,x_0)=1$. The expression for $Z_{\alpha_s} = \alpha_{s}/\alpha_{s 0}$ can be found from Eq. (\ref{One-LoopRunning}) and is written as

\begin{equation}
Z_{\alpha_{s}}\big(\alpha_s,\ln\Lambda/\mu\big)=\frac{\alpha_{s}}{\alpha_{s0}} = 1 +\frac{\alpha_{s}}{2\pi}\Big[3C_{2}\Big(\ln\frac{\Lambda}{\mu}+b_{11}\Big) -2N_{f}T(R) \Big(\ln\frac{\Lambda}{\mu}+b_{12}\Big)\Big] + O(\alpha_{s}^2).
\end{equation}

\noindent
Then the considered boundary condition gives $b_{11} = b_{12} = -x_0$.

Similarly, the second boundary condition in Eq. (\ref{Condition}) can be equivalently rewritten in the form  $\alpha_0^{-1}(\alpha,x_0) - \alpha^{-1} = 0$. Then Eq. (\ref{EquationForAlpha}) gives $d_1 = d_2 = -x_0$.\footnote{We will not present values for $d_{3i}$ and $g_{2i}$, because they do not enter the expression for $\widetilde\gamma(\alpha_s)$ and $\widetilde D(\alpha_s)$ in the considered approximation.} Imposing the first condition in Eq. (\ref{Condition}) on the function (\ref{Zed}) we obtain the equation $g_1=-x_0$. Therefore,

\begin{equation}\label{ConstantsEqualities}
g_1-b_{11} = 0;\qquad g_1-b_{12} = 0; \qquad d_2-b_{11}=0; \qquad d_2-b_{12}=0.
\end{equation}

\noindent
Consequently, the Adler $D$-function defined in terms of the renormalized coupling constant in the NSVZ-like scheme obtained with the higher covariant derivative regularization is

\begin{eqnarray}\label{ThreeLoopDRenormalizedFinal}
&& \widetilde{D}(\alpha_{s}) =\frac{3}{2}\sum\limits_{\alpha=1}^{N_{f}}q_{\alpha}^2\Bigg(\mbox{dim}(R) + \frac{\alpha_{s}}{\pi}\, \mbox{tr}\,C(R) +\frac{\alpha_s^2}{\pi^2}\Big[ - \frac{1}{2}\, \mbox{tr}\left(C(R)^2\right)  + \frac{3}{2}\, C_2\, \mbox{tr}\,C(R)\,  \Big(\ln a_\varphi + 1 \Big)\quad\nonumber\\
&& - N_{f} T(R)\,\mbox{tr}\,C(R)\, \Big( \ln a + 1 \Big)\Big] + O(\alpha_{s}^3)\Bigg).\vphantom{\frac{1}{2}}
\end{eqnarray}

\noindent
For the particular case of the $SU(N)$ gauge group and the matter superfields in the fundamental ($\phi$) and antifundamental ($\widetilde\phi$) representations we obtain

\begin{equation}
C_2 = N;\qquad T(R) = \frac{1}{2};\qquad C(R)_i{}^j = \frac{N^2-1}{2N} \delta_i^j;\qquad \mbox{dim}(R)=N,
\end{equation}

\noindent
so that in the NSVZ-like scheme the anomalous dimension $\left(\,\widetilde\gamma(\alpha_s)_i{}^j \equiv \widetilde\gamma(\alpha_s) \smash{\delta_i^j}\,\right)$ and the $D$-function take the form

\begin{eqnarray}
&& \widetilde\gamma(\alpha_s) = - \frac{\alpha_{s}}{\pi}\, \frac{N^2-1}{2N} +\frac{\alpha_s^2}{\pi^2}\Big[\, \frac{1}{2}\, \Big(\frac{N^2-1}{2N}\Big)^2 - \frac{3}{4}\, \big(N^2-1\big)  \Big(\ln a_\varphi + 1 \Big) + \frac{N_{f}}{4 N} \big(N^2-1\big)\qquad\vphantom{\Bigg)}\nonumber\\
&& \times \Big( \ln a + 1 \Big)\Big] + O(\alpha_{s}^3);\vphantom{\Bigg(}\\
&& \widetilde{D}(\alpha_{s}) =\frac{3}{2} N \sum\limits_{\alpha=1}^{N_{f}}q_{\alpha}^2\Bigg(1 + \frac{\alpha_{s}}{\pi}\, \frac{N^2-1}{2N} +\frac{\alpha_s^2}{\pi^2}\Big[ - \frac{1}{2}\, \Big(\frac{N^2-1}{2N}\Big)^2 + \frac{3}{4}\, \big(N^2-1\big)  \Big(\ln a_\varphi + 1 \Big) \qquad\nonumber\\
&& - \frac{N_{f}}{4 N} \big(N^2-1\big) \Big( \ln a + 1 \Big)\Big] + O(\alpha_{s}^3)\Bigg).
\end{eqnarray}

Comparing Eqs. (\ref{TwoLoopGammaBare}) with (\ref{TwoLoopGammaRenormalized}) and (\ref{ThreeLoopDBare}) with (\ref{ThreeLoopDRenormalized}) we conclude that Eq. (\ref{ConstantsEqualities}) ensures the identical equality between $\gamma$ and $\widetilde{\gamma}$ as well as between $D$ and $\widetilde{D}$ in the considered approximation. This exactly confirms the general arguments presented in the previous section. Moreover, it is evident now that in the subtraction scheme (\ref{Condition}) the RG functions defined in terms of the renormalized coupling constant satisfy the NSVZ-like relation

\begin{equation}\label{RenormalizedNSVZ}
\widetilde D(\alpha_{s})=\frac{3}{2}\sum\limits_{\alpha=1}^{N_{f}} q_{\alpha}^2\Big(\mbox{dim}(R) -\mbox{tr}\,\widetilde\gamma(\alpha_{s})\Big)
\end{equation}

\noindent
in the considered approximation.

\section{Conclusion}
\hspace*{\parindent}

In the three-loop approximation for ${\cal N}=1$ SQCD regularized by higher covariant derivatives we have verified the validity of the NSVZ-like equation (\ref{ShifmanFormula}) (and its more general version (\ref{WhatWeWantToProve})), which relates the Adler $D$-function to the anomalous dimension of the matter superfields defined in terms of the bare coupling constant. Moreover, we have obtained an explicit expression for the Adler function in the NSVZ-like scheme to the order $O(\alpha_s^2)$.

In full accordance with Ref. \cite{Shifman:2015doa} the three-loop $D$-function defined in terms of the bare coupling constant is given by integrals of double total derivatives. After integration with respect to the momentum of the matter loop they coincide with the integrals giving the two-loop anomalous dimension. Note that doing the calculations we use a version of the higher covariant derivative regularization that preserves the BRST invariance of the action. Although it leads to complicated integrals, for the higher derivative regulator (\ref{ExplicitRegulator}) the RG functions in the considered approximation can be explicitly calculated. Note that the RG functions defined in terms of the bare coupling constant satisfy the NSVZ-like relation independently of the renormalization prescription with the higher covariant derivative regularization. However, for the RG functions defined in terms of the renormalized coupling constant this relation is valid only in a certain subtraction scheme. In this paper we present the prescription which gives this scheme in all loops. It has also been verified by the explicit three-loop calculation, which confirmed its correctness in the considered approximation.

It would be also interesting to calculate $O(\alpha_s^2)$ contribution to the Adler $D$-function in the $\overline{\mbox{DR}}$-scheme. Having in mind the results of \cite{Jack:1996vg,Jack:1996cn}, one may suggest that the NSVZ-like equation (\ref{RenormalizedNSVZ}) is not satisfied in the $\overline{\mbox{DR}}$-scheme, and a specially tuned finite renormalization is needed to obtain the RG functions for which it is valid.

\section*{Acknowledgements}
\hspace*{\parindent}

We are deeply grateful to M.A.Shifman for the interest to this work. We are also grateful to A.A.Vladimirov for the discussion of several topics related to this paper and to E.A.Ivanov for the valuable discussion and comments.

The work of A.L.K. and A.E.K. was supported by the grant of the Foundation "BASIS" No. 17-11-120.

\appendix

\section{Superdiagrams contributing to the $D$-function}
\hspace*{\parindent}\label{AppendixGraphs}

Let us present expressions for contributions of the supergraphs in Fig. \ref{Figure_Adler} to the expression

\begin{equation}
\frac{d}{d\ln\Lambda}\Big(d^{-1} - \alpha_0^{-1}\Big)\Big|_{p=0} = \frac{2}{3\pi} D(\alpha_{s0}).
\end{equation}

\noindent
Each of them corresponds to the sum of all superdiagrams obtained from a considered supergraph by attaching two external lines of the Abelian gauge superfield. All these expressions are written in the Euclidean space after the Wick rotation. Note that, for convenience, some contributions are summed.

\begin{eqnarray}\label{Graph1}
&& \frac{d}{d\ln\Lambda} \mbox{graph} (1) = 4\pi\sum\limits_{\alpha=1}^{N_{f}}q_{\alpha}^{2}\, \mbox{tr}\,C(R)\frac{d}{d\ln\Lambda}\int\frac{d^4q}{(2\pi)^4}\frac{d^4k}{(2\pi)^4}\frac{1}{k^2R_{k}} \frac{\partial}{\partial q^{\mu}}\frac{\partial}{\partial q_{\mu}} \Big(\frac{g_0^2}{q^2 (q+k)^2}\nonumber\\
&&\qquad -\frac{2(g_0^2-g^2)}{k^2q^2} - \frac{g_{0}^2}{(q^2+M^2)((q+k)^2+M^2)}+\frac{2(g_{0}^2-g^2)}{k^2(q^2+M^2)}\Big);\\
&& \frac{d}{d\ln\Lambda} \mbox{graph} (2) = 8\pi\sum\limits_{\alpha=1}^{N_{f}}q_{\alpha}^{2}\, \mbox{tr}\,C(R) \frac{d}{d\ln\Lambda}\int\frac{d^4q}{(2\pi)^4}\frac{d^4k}{(2\pi)^4}\frac{g_{0}^2-g^{2}}{k^4R_{k}}\frac{\partial}{\partial q^{\mu}}\frac{\partial}{\partial q_{\mu}} \Big(\frac{1}{q^2} - \frac{1}{q^2+M^2}\Big);\nonumber\\
&&\vphantom{1}\\
&& \frac{d}{d\ln\Lambda} \mbox{graph} (3) = 8\pi\sum\limits_{\alpha=1}^{N_{f}}q_{\alpha}^{2}\, \mbox{tr}\left(C(R)^2\right) \frac{d}{d\ln\Lambda}\int\frac{d^4q}{(2\pi)^4}\frac{d^4k}{(2\pi)^4}\frac{d^4l}{(2\pi)^4}\frac{g_{0}^4}{k^2R_{k}l^2R_{l}} \frac{\partial}{\partial q^{\mu}}\frac{\partial}{\partial q_{\mu}}\nonumber\\
&&\qquad \times \Big(\frac{1}{q^2 (q+k)^2 (q+l)^2} - \frac{q^2-M^2}{(q^2+M^2)^2((q+k)^2+M^2)((q+l)^2+M^2)}\Big);\\
&& \frac{d}{d\ln\Lambda} \mbox{graph} (4) = 4\pi\sum\limits_{\alpha=1}^{N_{f}}q_{\alpha}^{2}\, \mbox{tr} \Big(C(R)^2-\frac{C_{2}C(R)}{2}\Big) \frac{d}{d\ln\Lambda}\int\frac{d^4q}{(2\pi)^4}\frac{d^4k}{(2\pi)^4}\frac{d^4l}{(2\pi)^4}\frac{g_{0}^4}{k^2R_{k}l^2R_{l}}\nonumber\\
&&\qquad \times\frac{\partial}{\partial q^{\mu}}\frac{\partial}{\partial q_{\mu}}\Big(\frac{(k+l+2q)^2}{q^2 (q+k)^2 (q+l)^2 (q+k+l)^2} - \frac{(k+l+2q)^2+2M^2}{(q^2+M^2)((q+k)^2+M^2)((q+l)^2+M^2)}\nonumber\\
&&\qquad \times \frac{1}{((q+k+l)^2+M^2)}\Big);\\
&& \frac{d}{d\ln\Lambda} \mbox{graph} (5)= -16\pi\sum\limits_{\alpha=1}^{N_{f}}q_{\alpha}^{2}\, \mbox{tr}\Big(C(R)^2-\frac{C_2 C(R)}{4}\Big) \frac{d}{d\ln\Lambda}\int\frac{d^4q}{(2\pi)^4}\frac{d^4k}{(2\pi)^4}\frac{d^4l}{(2\pi)^4} \frac{g_0^4}{k^2R_{k}l^2R_{l}}\nonumber\\
&&\qquad \times \frac{\partial}{\partial q^{\mu}}\frac{\partial}{\partial q_{\mu}}\Big(\frac{1}{q^2 (q+k)^2 (q+l)^2} - \frac{1}{(q^2+M^2)((q+k)^2+M^2)((q+l)^2+M^2)}\Big);\\
&& \frac{d}{d\ln\Lambda} \mbox{graph} (6) = -8\pi\sum\limits_{\alpha=1}^{N_{f}}q_{\alpha}^{2} C_{2} \mbox{tr} \,C(R) \frac{d}{d\ln\Lambda} \int\frac{d^4q}{(2\pi)^4} \frac{d^4k}{(2\pi)^4} \frac{d^4l}{(2\pi)^4} \frac{g_{0}^4}{l^2R_{l}k^2R_{k}(l-k)^2} \frac{\partial}{\partial q^{\mu}}\frac{\partial}{\partial q_{\mu}}\nonumber\\
&&\qquad \times\Bigg[\frac{(q+l)_\mu (q+k)^\mu}{q^2 (q+l)^2 (q+k)^2} - \frac{(q+l)_\mu (q+k)^\mu +M^{2}}{(q^2+M^2)((q+l)^2+M^{2})((q+k)^2+M^2)} +\frac{2}{R_{k-l}} \frac{R_{l}-R_{k}}{l^2-k^2} \nonumber\\
&&\qquad \times \Bigg(\frac{q^2 (q+k)_\mu l^\mu + l^2 (q+k)_\mu q^\mu}{q^2 (q+l)^2 (q+k)^2} - \frac{q^2 (q+k)_\mu l^\mu + l^2 (q+k)_\mu q^\mu + M^2(q+l+k)_\mu l^\mu}{(q^2+M^2)((q+l)^2+M^2)((q+k)^2+M^2)}\Bigg)\Bigg];\nonumber\\
&& \vphantom{1}\\
&& \frac{d}{d\ln\Lambda}\Big(\mbox{graph}(7) + \mbox{graph}(8) + \mbox{graph}(9)\Big)=-8\pi\sum\limits_{\alpha=1}^{N_{f}}q_{\alpha}^{2}\, N_{f}T(R)\mbox{ tr}\,C(R)\frac{d}{d\ln\Lambda}\int\frac{d^4q}{(2\pi)^4}\nonumber\\
&&\qquad \times \int \frac{d^4k}{(2\pi)^4} \frac{d^4l}{(2\pi)^4}\frac{g_{0}^4}{k^2 R_{k}^2} \frac{\partial}{\partial q^{\mu}}\frac{\partial}{\partial q_{\mu}} \Bigg[ \Big(\frac{1}{q^2 (q+k)^2} - \frac{1}{\big(q^2+M^2\big)\big((q+k)^2 +M^2\big)}\Big) \Big(\frac{1}{l^2 (l+k)^2}\nonumber\\
&&\qquad - \frac{1}{\big(l^2+M^2\big)\big((l+k)^2 +M^2\big)}\Big)\Bigg];\\
&& \frac{d}{d\ln\Lambda}\Big(\mbox{graph}(10)+\mbox{graph}(11)+\dots +\mbox{graph}(21)\Big)=-8\pi\sum\limits_{\alpha=1}^{N_{f}}q_{\alpha}^2 C_{2} \mbox{tr} \,C(R)\frac{d}{d\ln\Lambda}\int\frac{d^4q}{(2\pi)^4}\nonumber\\
&& \times \int\frac{d^4k}{(2\pi)^4}\frac{g_{0}^4}{k^2R_{k}^2}\frac{\partial}{\partial q^{\mu}}\frac{\partial}{\partial q_{\mu}}\Big(\frac{f(k/\Lambda)}{q^2 (q+k)^2} - \frac{f(k/\Lambda)}{(q^2+M^2)((q+k)^2+M^2)}\Big),
\end{eqnarray}

\noindent
where the function $f(k/\Lambda)$, related to the one-loop polarization operator, has the form

\begin{eqnarray}\label{Polarization}
&&\hspace*{-3mm} f(k/\Lambda) = - \int\frac{d^4l}{(2\pi)^4} \Biggl[\frac{3}{2} \left(\frac{1}{l^2(l+k)^2}-\frac{1}{(l^2+M_{\varphi}^2)((l+k)^2+M_{\varphi}^2)}\right) -\frac{R_{l}-R_{k}}{R_{l}l^2}\left(\frac{1}{(l+k)^2}\right.\nonumber\\
&&\hspace*{-3mm} \left.-\frac{1}{l^2-k^2}\right) -\frac{2}{R_l \big((l+k)^2-l^2\big)}\left(\frac{R_{l+k}-R_{l}}{(l+k)^2-l^2}-\frac{R'_{l}}{\Lambda^2}\right) + \frac{1}{R_{l}R_{l+k}} \left(\frac{R_{l+k}-R_{l}}{(l+k)^2-l^2}\right)^2  \nonumber\\
&&\hspace*{-3mm} + \frac{2 R_{k}k^2}{l^2(l+k)^2R_{l}R_{l+k}} \left(\frac{R_{l+k}-R_{k}}{(l+k)^2-k^2}\right) + \frac{l_\mu k^\mu R_{k}}{l^2R_{l}(l+k)^2R_{l+k}}\left(\frac{R_{l+k}-R_{l}}{(l+k)^2-l^2}\right) - \frac{2 l_\mu k^\mu}{l^2R_{l}R_{l+k}}\nonumber\\
&&\hspace*{-3mm} \times\left(\frac{R_{l+k}-R_{k}}{(l+k)^2-k^2}\right) \left(\frac{R_{l+k}-R_{l}}{(l+k)^2-l^2}\right) + \frac{2 k^2}{(l+k)^2R_{l}R_{l+k}} \left(\frac{R_{l}-R_{k}}{l^2-k^2}\right)^2 +\frac{k^2 l_\mu (l+k)^\mu}{l^2(l+k)^2R_{l}R_{l+k}}\nonumber\\
&&\hspace*{-3mm} \times \left(\frac{R_{l}-R_{k}}{l^2-k^2}\right) \left(\frac{R_{l+k}-R_{k}}{(l+k)^2-k^2}\right) -\frac{2}{(l+k)^2-k^2}\left(\frac{R_{l+k}-R_{k}}{(l+k)^2-k^2} -\frac{R'_{k}}{\Lambda^2}\right)\frac{k^2}{l^2R_{l}} +\frac{2 l_\mu k^\mu}{l^2R_{l}}\nonumber\\
&&\hspace*{-3mm} \times \left(\frac{R_{l}}{\left(l^2-(l+k)^2\right) \left(l^2-k^2\right)}+\frac{R_{l+k}}{\left((l+k)^2-l^2\right) \left((l+k)^2-k^2\right)} +\frac{R_{k}}{\left(k^2-l^2\right)\left(k^2-(l+k)^2\right)}\right)\nonumber\\
&&\hspace*{-3mm} + \frac{1}{2\big((l+k)^2-l^2\big)}\left(\frac{2 R_{l+k} R'_{l+k} (l+k)^2}{\Lambda^2\big((l+k)^2 R_{l+k}^2+M_\varphi^2\big)} -\frac{2 R_l R'_l l^2}{\Lambda^2\big(l^2 R_l^2 + M_\varphi^2\big)} - \frac{1}{(l+k)^2 + M_\varphi^2} + \frac{1}{l^2+M_\varphi^2}\right.\nonumber\\
&&\hspace*{-3mm} \left. +\frac{R_{l+k}^2}{(l+k)^2 R_{l+k}^2 + M_\varphi^2} - \frac{R_l^2}{l^2 R_l^2+M_\varphi^2} \right)\Biggr];
\end{eqnarray}

\begin{equation}
\frac{d}{d\ln\Lambda} \Big(\mbox{graph}(22)\Big) = \frac{d}{d\ln\Lambda} \Big(\mbox{graph}(23)\Big) = 0.\qquad\qquad\qquad\qquad\qquad\qquad\qquad\quad
\end{equation}

\section{Supergraphs contributing to the two-point Green function of the matter superfields}
\hspace*{\parindent}\label{AppendixMatter}

For the sake of completeness, in this section we present expressions for contributions of superdiagrams presented in Fig. \ref{Figure_Gamma} to the function $G_{i}{}^{j}(\alpha_{s0},\Lambda/q)$. They are written in the Euclidean space after the Wick rotation. For convenience, certain groups of diagrams are summed.

\begin{eqnarray}
&&\hspace*{-6mm} \big(\Delta G^{(1)}\big)_i{}^j =-C(R)_i{}^j \int \frac{d^4k}{(2\pi)^4} \frac{2}{k^2R_{k}} \Big(\frac{g_{0}^2}{(q+k)^2}-\frac{(g_{0}^2-g^2)((q+k)^2+q^2)}{k^2(q+k)^2}\Big);\qquad\\
&&\hspace*{-6mm} \big(\Delta G^{(2)}\big)_i{}^j =-C(R)_i{}^j\int\frac{d^4k}{(2\pi)^4}\frac{2(g_{0}^2-g^2)}{k^4R_{k}};\\
&&\hspace*{-6mm} \big(\Delta G^{(3.2)}\big)_i{}^j =-\big(C(R)^2\big)_i{}^j \int\frac{d^4k}{(2\pi)^4}\frac{d^4l}{(2\pi)^4}\frac{4g_{0}^4}{k^2R_{k}l^2R_{l}(q+k)^2(q+k+l)^2};\\
&&\hspace*{-6mm} \big(\Delta G^{(4)}\big)_i{}^j =-\Big(\big(C(R)^2\big)_i{}^j - \frac{C_{2} C(R)_i{}^j}{2}\Big) \int\frac{d^4k}{(2\pi)^4}\frac{d^4l}{(2\pi)^4}\frac{4g_{0}^4(2q+k+l)^2}{k^2R_{k}l^2R_{l}(q+k)^2(q+l)^2(q+k+l)^2};\nonumber\\
&&\vphantom{1}\\
&&\hspace*{-6mm} \big(\Delta G^{(5.1)}\big)_i{}^j =\Big(\big(C(R)^2\big)_i{}^j - \frac{C_{2}C(R)_i{}^j}{4}\Big) \int\frac{d^4k}{(2\pi)^4}\frac{d^4l}{(2\pi)^4}\frac{8g_{0}^4}{k^2R_{k}l^2R_{l}(q+k)^2(q+k+l)^2};\\
&&\hspace*{-6mm} \big(\Delta G^{(5.2)}\big)_i{}^j =\Big(\big(C(R)^2\big)_i{}^j - \frac{C_{2}C(R)_i{}^j}{4}\Big) \int\frac{d^4k}{(2\pi)^4}\frac{d^4l}{(2\pi)^4}\frac{4g_{0}^4}{k^2R_{k}l^2R_{l}(q+k)^2(q+l)^2};\\
&&\hspace*{-6mm} \big(\Delta G^{(6)}\big)_i{}^j = C_{2} C(R)_i{}^j \int\frac{d^4k}{(2\pi)^4} \frac{d^4l}{(2\pi)^4} \frac{2g_{0}^4\left(R_{l}(q+l)_\mu q^\mu + R_{k} (q-k)_\mu q^\mu + R_{l+k} (q+l)_\mu (q-k)^\mu\vphantom{R_a^b}\right)}{k^2R_{k}l^2R_{l}(l+k)^2R_{l+k}(q+l)^2(q-k)^2}
\nonumber\\
&& + C_{2}C(R)_i{}^j \int\frac{d^4k}{(2\pi)^4}\frac{d^4l}{(2\pi)^4}\frac{4g_{0}^4}{k^2R_{k}l^2R_{l}(l+k)^2R_{l+k}(q-k)^2(q+l)^2}\Bigg[\frac{R_{l}-R_{k}}{l^2-k^2}\Big((q+l)_\mu q^\mu\, \nonumber\\
&& \times (q-k)_\nu l^\nu + (q+l)_\mu l^\mu\, q_\nu (q-k)^\nu - (q+l)_\mu (q-k)^\mu\, q_\nu l^\nu\Big)
+\frac{R_{l+k}-R_{k}}{(l+k)^2-k^2}\Big((q+l)_\mu\nonumber\\
&&\times (q-k)^\mu\, q_\nu k^\nu + (q+l)_\mu k^\mu\, (q-k)_\nu q^\nu - (q+l)_\mu q^\mu \,(q-k)_\nu k^\nu \Big) -\frac{R_{l+k}-R_{l}}{(l+k)^2-l^2}\Big(q_\mu (l+k)^\mu\,\nonumber\\
&&\times (q-k)_\nu (q+l)^\nu + q_\mu (q+l)^\mu\, (l+k)_\nu (q-k)^\nu - q_\mu (q-k)^\mu\, (l+k)_\nu (q+l)^\nu \Big)\Bigg];\\
&&\hspace*{-6mm} \big(\Delta G^{(7.1)}\big)_i{}^j + \big(\Delta G^{(7.2)}\big)_i{}^j + \big(\Delta G^{(8)}\big)_i{}^j + \big(\Delta G^{(9)}\big)_i{}^j = N_{f} T(R)\, C(R)_i{}^j \int\frac{d^4k}{(2\pi)^4}\frac{d^4l}{(2\pi)^4}\frac{4g_{0}^4}{k^4 R_{k}^2}\nonumber\\
&& \times \frac{k^2-q^2}{(q+k)^2}\,\Big(\frac{1}{l^2 (l+k)^2} - \frac{1}{\big(l^2 + M^2\big)\big((l+k)^2 + M^2\big)}\Big);\\
&&\hspace*{-6mm} \big(\Delta G^{(10)}\big)_i{}^j +\cdots + \big(\Delta G^{(21)}\big)_i{}^j = C_{2} C(R)_i{}^j \int\frac{d^4k}{(2\pi)^4} \frac{4g_{0}^4}{k^2R_{k}^2}f(k/\Lambda)\Big(\frac{1}{(q+k)^2}-\frac{q^2}{k^2(q+k)^2}\Big);\\
&&\hspace*{-6mm} \big(\Delta G^{(22)}\big)_i{}^j = \big(\Delta G^{(23)}\big)_i{}^j = 0,\vphantom{\frac{1}{2}}
\end{eqnarray}

\noindent
where $f(k/\Lambda)$ is the same function as earlier given by Eq. (\ref{Polarization}).

\section{Calculation of the integrals}
\hspace*{\parindent}\label{AppendixIntegrals}

Let us calculate the anomalous dimension defined in terms of the bare coupling constant given by Eq. (\ref{FinalIntegralGamma}) for the regulator function (\ref{ExplicitRegulator}). It is important that the differentiation with respect to $\ln\Lambda$ should be done at a fixed value of the renormalized coupling constant $g$ (or, equivalently, $\alpha_s = g^2/4\pi$) before the integrations. That is why, first, we rewrite the expression for the anomalous dimension in terms of $\alpha_s$ using Eq. (\ref{One-LoopRunning}). In the considered approximation this gives

\begin{eqnarray}
&&\hspace*{-3mm} \gamma_i{}^j(\alpha_{s0}) = - 8\pi\alpha_s C(R)_i{}^j \int\frac{d^4k}{(2\pi)^4} \frac{d}{d\ln\Lambda} \frac{1}{k^4R_{k}} + 64\pi^2\alpha_s^2 C_{2} C(R)_i{}^j \int\frac{d^4k}{(2\pi)^4} \frac{1}{k^4} \frac{d}{d\ln\Lambda}  \nonumber\\
&&\hspace*{-3mm} \times \Big[\frac{f(k/\Lambda)}{R_k^2} + \frac{3}{16\pi^2 R_k}\Big(\ln \frac{\Lambda}{\mu}+ b_{11}\Big) \Big] + 64\pi^2 \alpha_s^2 N_{f} T(R) C(R)_i{}^j \int\frac{d^4k}{(2\pi)^4} \frac{1}{k^4} \frac{d}{d\ln\Lambda} \Big[\frac{1}{R_k^2} \nonumber\\
&&\hspace*{-3mm} \times \int \frac{d^4l}{(2\pi)^4} \Big(\frac{1}{l^2 (l+k)^2} - \frac{1}{\big(l^2+M^2\big)\big((l+k)^2+M^2\big)}\Big) - \frac{1}{8\pi^2 R_k}\Big(\ln \frac{\Lambda}{\mu}+ b_{12}\Big)\Big] - 32\pi^2 \nonumber\\
&&\hspace*{-3mm} \times \alpha_s^2 \left(C(R)^2\right)_i{}^j \int\frac{d^4k}{(2\pi)^4}\frac{d^4l}{(2\pi)^4} \frac{d}{d\ln\Lambda} \frac{1}{k^2R_{k}l^2R_{l}}\Big(\frac{1}{l^2k^2} -\frac{2}{l^2(l+k)^2}\Big) + O\left(\alpha_s^3\right).
\end{eqnarray}

\noindent
Some integrals in this equation can be calculated using the results of Refs. \cite{Soloshenko:2002np,Soloshenko:2003sx} for the simplest regulator $R(y)=1+y^n$, where $n$ is a positive integer,

\begin{eqnarray}
&& \int \frac{d^4k}{(2\pi)^4}\, \frac{d}{d\ln\Lambda} \frac{1}{k^4 R_k} = \frac{1}{8\pi^2};\\
&& \int \frac{d^4k}{(2\pi)^4}\, \frac{d^4l}{(2\pi)^4}\,\frac{d}{d\ln\Lambda} \frac{1}{k^2 R_k\,l^2 R_l} \Big(\frac{1}{l^2 k^2} - \frac{2}{l^2 (l+k)^2} \Big) = - \frac{1}{64\pi^4};\\
&& \int \frac{d^4k}{(2\pi)^4}\, \frac{1}{k^4} \frac{d}{d\ln\Lambda} \Big[\frac{1}{R_k^2} \int \frac{d^4l}{(2\pi)^4}\, \Big(\frac{1}{l^2 (l+k)^2} - \frac{1}{\big(l^2+M^2\big)\big((l+k)^2+M^2\big)}\Big)\nonumber\\
&&\qquad\qquad\qquad\ \ \, - \frac{1}{8\pi^2 R_k}\Big(\ln \frac{\Lambda}{\mu}+ b_{12}\Big)\Big] = \frac{1}{64\pi^4}\Big(- \ln \frac{\Lambda}{\mu} - b_{12} + \ln a + 1 \Big),\qquad
\end{eqnarray}

\noindent
where $a = M/\Lambda$.

Thus, it is necessary to calculate only the integral

\begin{equation}
\int\frac{d^4k}{(2\pi)^4} \frac{1}{k^4} \frac{d}{d\ln\Lambda} \Big[\frac{f(k/\Lambda)}{R_k^2} + \frac{3}{16\pi^2 R_k}\Big(\ln \frac{\Lambda}{\mu}+ b_{11}\Big) \Big].
\end{equation}

\noindent
Let us split the function $f(k/\Lambda)$ given by Eq. (\ref{Polarization}) into two parts,

\begin{equation}
f(k/\Lambda) \equiv f_1(k/\Lambda) + f_2(k/\Lambda),
\end{equation}

\noindent
where

\begin{equation}
f_1(k/\Lambda) \equiv  - \frac{3}{2} \int\frac{d^4l}{(2\pi)^4} \Big(\frac{1}{l^2(l+k)^2}-\frac{1}{(l^2+M_{\varphi}^2)((l+k)^2+M_{\varphi}^2)}\Big)
\end{equation}

\noindent
and the function $f_2(k/\Lambda)$ includes all remaining terms in the expression (\ref{Polarization}).

The integral containing the function $f_1(k/\Lambda)$ can again be calculated using the results of Refs. \cite{Soloshenko:2002np,Soloshenko:2003sx},

\begin{equation}
\int\frac{d^4k}{(2\pi)^4} \frac{1}{k^4} \frac{d}{d\ln\Lambda} \Big[\frac{f_1(k/\Lambda)}{R_k^2} + \frac{3}{16\pi^2 R_k}\Big(\ln \frac{\Lambda}{\mu}+ b_{11}\Big) \Big] = -\frac{3}{128\pi^4}\Big(- \ln \frac{\Lambda}{\mu} - b_{11} + \ln a_\varphi + 1 \Big),
\end{equation}

\noindent
where $a_\varphi \equiv M_\varphi/\Lambda$.

Therefore, we should calculate only the remaining integral containing the function $f_2(k/\Lambda)$,

\begin{equation}
I\equiv \int\frac{d^4k}{(2\pi)^4} \frac{1}{k^4} \frac{d}{d\ln\Lambda} \frac{f_2(k/\Lambda)}{R_k^2}.
\end{equation}

\noindent
First, we note that the function $f_2(k/\Lambda)/R_k^2$ depends only on the ratio $k/\Lambda$. Consequently, the considered integral can be rewritten as

\begin{equation}
I = - \int\frac{d^4k}{(2\pi)^4} \frac{1}{k^4} \frac{d}{d\ln k} \frac{f_2(k/\Lambda)}{R_k^2} =  - 2 \int\frac{d^4k}{(2\pi)^4} \frac{1}{k^2} \frac{d}{dk^2} \frac{f_2(k/\Lambda)}{R_k^2}.
\end{equation}

\noindent
Calculating this integral in the four-dimensional spherical coordinates, we obtain

\begin{equation}
I = - \frac{1}{8\pi^2} \frac{f_2(k/\Lambda)}{R_k^2}\Bigg|_0^\infty = \frac{1}{8\pi^2} f_2(0),
\end{equation}

\noindent
where we take into account that the function $f_2(k/\Lambda)/R_k^2$ vanishes in the limit $k\to \infty$. In the limit $k\to 0$ the function $f_2(k/\Lambda)$ can be written as an integral of double total derivatives,

\begin{eqnarray}\label{F2_At_Zero}
&& f_2(0) = \int \frac{d^4l}{(2\pi)^4} \frac{d}{dl^2} \Big(\frac{1}{2(l^2 + M_\varphi^2)} - \frac{R_l^2}{2(l^2 R_l^2 + M_\varphi^2)} + \frac{M_\varphi^2 R_l'}{\Lambda^2 R_l\big(l^2 R_l^2 + M_\varphi^2\big)}\Big)\qquad\nonumber\\
&& = -\frac{1}{8} \int\frac{d^4l}{(2\pi)^4}\,\frac{\partial}{\partial l^\mu} \frac{\partial}{\partial l_\mu} \Bigg[\frac{1}{l^2} \ln \frac{l^2 R_l^2 + M_\varphi^2}{R_l^2(l^2 + M_\varphi^2)}\Bigg] = 0.
\end{eqnarray}

\noindent
The last equality follows from the fact that the expression in the square brackets is not singular in the limit $l\to 0$. (Note that Eq. (\ref{F2_At_Zero}) is valid for an arbitrary function $R(x)$ rapidly growing at infinity, such that $R(0)=1$.) Therefore, we obtain

\begin{equation}
I = 0.
\end{equation}

Collecting the results for all integrals, the anomalous dimension defined in terms of the bare coupling constant can be presented as

\begin{eqnarray}
&&\hspace*{-5mm} \gamma_i{}^j(\alpha_{s0}) = - \frac{\alpha_s}{\pi}\, C(R)_i{}^j - \frac{3\alpha_s^2}{2\pi^2}\, C_{2} C(R)_i{}^j
\Big(- \ln \frac{\Lambda}{\mu} - b_{11} + \ln a_\varphi + 1 \Big) + \frac{\alpha_s^2}{\pi^2}\, N_{f} T(R) C(R)_i{}^j\nonumber\\
&&\hspace*{-5mm} \times \Big(- \ln \frac{\Lambda}{\mu} - b_{12} + \ln a + 1 \Big) + \frac{\alpha_s^2}{2\pi^2}\, \left(C(R)^2\right)_i{}^j + O\left(\alpha_s^3\right).
\end{eqnarray}

\noindent
Rewriting the right hand side in terms of the bare coupling constant $\alpha_{s0}$ we finally obtain

\begin{eqnarray}
&& \gamma_i{}^j(\alpha_{s0}) = - \frac{\alpha_{s0}}{\pi}\, C(R)_i{}^j - \frac{3\alpha_{s0}^2}{2\pi^2}\, C_{2} C(R)_i{}^j
\big( \ln a_\varphi + 1 \big) + \frac{\alpha_{s0}^2}{\pi^2}\, N_{f} T(R) C(R)_i{}^j \big( \ln a + 1 \big)\nonumber\\
&& + \frac{\alpha_{s0}^2}{2\pi^2}\, \left(C(R)^2\right)_i{}^j + O\left(\alpha_{s0}^3\right).
\end{eqnarray}


\begin{thebibliography}{100}

%\cite{Adler:1974gd}
\bibitem{Adler:1974gd}
  S.~L.~Adler,
  %``Some Simple Vacuum Polarization Phenomenology: e+ e- ---> Hadrons: The mu - Mesic Atom x-Ray Discrepancy and (g-2) of the Muon,''
  Phys.\ Rev.\ D {\bf 10} (1974) 3714.
  %%CITATION = PHRVA,D10,3714;%%

%\cite{Jegerlehner:2017lbd}
\bibitem{Jegerlehner:2017lbd}
  F.~Jegerlehner,
  ``Muon g-2 Theory: the Hadronic Part,''
  arXiv:1705.00263 [hep-ph].
  %%CITATION = ARXIV:1705.00263;%%

%\cite{Eidelman:1998vc}
\bibitem{Eidelman:1998vc}
  S.~Eidelman, F.~Jegerlehner, A.~L.~Kataev and O.~Veretin,
  %``Testing nonperturbative strong interaction effects via the Adler function,''
  Phys.\ Lett.\ B {\bf 454} (1999) 369.
  %doi:10.1016/S0370-2693(99)00389-5
  %[hep-ph/9812521].
 %%CITATION = doi:10.1016/S0370-2693(99)00389-5;%%

%\cite{Chetyrkin:1979bj}
\bibitem{Chetyrkin:1979bj}
  K.~G.~Chetyrkin, A.~L.~Kataev and F.~V.~Tkachov,
  %``Higher Order Corrections to Sigma-t (e+ e- ---> Hadrons) in Quantum Chromodynamics,''
  Phys.\ Lett.\  {\bf 85B} (1979) 277.
  %doi:10.1016/0370-2693(79)90596-3
  %%CITATION = doi:10.1016/0370-2693(79)90596-3;%%

%\cite{Celmaster:1979xr}
\bibitem{Celmaster:1979xr}
  W.~Celmaster and R.~J.~Gonsalves,
  %``An Analytic Calculation of Higher Order Quantum Chromodynamic Corrections in e+ e- Annihilation,''
  Phys.\ Rev.\ Lett.\  {\bf 44} (1980) 560.
  %doi:10.1103/PhysRevLett.44.560
  %%CITATION = doi:10.1103/PhysRevLett.44.560;%%

%\cite{Dine:1979qh}
\bibitem{Dine:1979qh}
  M.~Dine and J.~R.~Sapirstein,
  %``Higher Order QCD Corrections in e+ e- Annihilation,''
  Phys.\ Rev.\ Lett.\  {\bf 43} (1979) 668.
  %doi:10.1103/PhysRevLett.43.668
  %%CITATION = doi:10.1103/PhysRevLett.43.668;%%

%\cite{Gorishnii:1990vf}
\bibitem{Gorishnii:1990vf}
  S.~G.~Gorishny, A.~L.~Kataev and S.~A.~Larin,
  %``The $O(\alpha^{3}_{s})$-corrections to $\sigma_{tot}(e^{+}e^{-}\rightarrow hadrons)$ and $\Gamma(\tau^{-} \rightarrow \nu_{\tau} + hadrons)$ in QCD,''
  Phys.\ Lett.\ B {\bf 259} (1991) 144.
  %doi:10.1016/0370-2693(91)90149-K
  %%CITATION = doi:10.1016/0370-2693(91)90149-K;%%

%\cite{Surguladze:1990tg}
\bibitem{Surguladze:1990tg}
  L.~R.~Surguladze and M.~A.~Samuel,
  %``Total hadronic cross-section in e+ e- annihilation at the four loop level of perturbative QCD,''
  Phys.\ Rev.\ Lett.\  {\bf 66} (1991) 560
   Erratum: [Phys.\ Rev.\ Lett.\  {\bf 66} (1991) 2416].
  %doi:10.1103/PhysRevLett.66.560
  %%CITATION = doi:10.1103/PhysRevLett.66.560;%%

%\cite{Chetyrkin:1996ez}
\bibitem{Chetyrkin:1996ez}
  K.~G.~Chetyrkin,
  %``Corrections of order alpha-s**3 to R(had) in pQCD with light gluinos,''
  Phys.\ Lett.\ B {\bf 391} (1997) 402.
  %doi:10.1016/S0370-2693(96)01478-5
  %[hep-ph/9608480].

%\cite{Baikov:2008jh}
\bibitem{Baikov:2008jh}
  P.~A.~Baikov, K.~G.~Chetyrkin and J.~H.~Kuhn,
  %``Order alpha**4(s) QCD Corrections to Z and tau Decays,''
  Phys.\ Rev.\ Lett.\  {\bf 101} (2008) 012002.
  %[arXiv:0801.1821 [hep-ph]].
  %%CITATION = ARXIV:0801.1821;%%

%\cite{Baikov:2012zn}
\bibitem{Baikov:2012zn}
  P.~A.~Baikov, K.~G.~Chetyrkin, J.~H.~Kuhn and J.~Rittinger,
  %``Adler Function, Sum Rules and Crewther Relation of Order O(alpha_s^4): the Singlet Case,''
  Phys.\ Lett.\ B {\bf 714} (2012) 62.
  %doi:10.1016/j.physletb.2012.06.052
  %[arXiv:1206.1288 [hep-ph]].
  %%CITATION = doi:10.1016/j.physletb.2012.06.052;%%

%\cite{Chetyrkin:1983qc}
\bibitem{Chetyrkin:1983qc}
  K.~G.~Chetyrkin, S.~G.~Gorishny, A.~L.~Kataev, S.~A.~Larin and F.~V.~Tkachov,
  %``SCALAR QUARKS: HIGHER CORRECTIONS TO sigma-t (e+ e- ---> HADRONS),''
  Phys.\ Lett.\  {\bf 116B} (1981) 455.
  %doi:10.1016/0370-2693(82)90167-8
  %%CITATION = doi:10.1016/0370-2693(82)90167-8;%%

%\cite{Mihaila:2013wma}
\bibitem{Mihaila:2013wma}
  L.~Mihaila,
  %``Precision Calculations in Supersymmetric Theories,''
  Adv.\ High Energy Phys.\  {\bf 2013} (2013) 607807.
  %doi:10.1155/2013/607807
  %[arXiv:1310.6178 [hep-ph]].
  %%CITATION = doi:10.1155/2013/607807;%%

%\cite{Kataev:1983at}
\bibitem{Kataev:1983at}
  A.~L.~Kataev and A.~A.~Pivovarov,
  %``PERTURBATIVE CORRECTIONS TO sigma-t (e+ e- ---> HADRONS) IN SUPERSYMMETRIC QCD,''
  JETP Lett.\  {\bf 38} (1983) 369
   [Pisma Zh.\ Eksp.\ Teor.\ Fiz.\  {\bf 38} (1983) 309].
  %%CITATION = JTPLA,38,369;%%

%\cite{Altarelli:1983pr}
\bibitem{Altarelli:1983pr}
  G.~Altarelli, B.~Mele and R.~Petronzio,
  %``Broken Supersymmetric {QCD} and $e^+ e^-$ Hadronic Cross-sections,''
  Phys.\ Lett.\  {\bf 129B} (1983) 456.
  %doi:10.1016/0370-2693(83)90139-9
  %%CITATION = doi:10.1016/0370-2693(83)90139-9;%%

%\cite{Shifman:2014cya}
\bibitem{Shifman:2014cya}
  M.~Shifman and K.~Stepanyantz,
  %``Exact Adler Function in Supersymmetric QCD,''
  Phys.\ Rev.\ Lett.\  {\bf 114} (2015) 051601.
  %[arXiv:1412.3382 [hep-th]].
  %%CITATION = ARXIV:1412.3382;%%

%\cite{Shifman:2015doa}
\bibitem{Shifman:2015doa}
  M.~Shifman and K.~V.~Stepanyantz,
  %``Derivation of the exact expression for the D function in N=1 SQCD,''
  Phys.\ Rev.\ D {\bf 91} (2015) 105008.
  %[arXiv:1502.06655 [hep-th]].

%\cite{Slavnov:1971aw}
\bibitem{Slavnov:1971aw}
  A.~A.~Slavnov,
  %``Invariant regularization of nonlinear chiral theories,''
  Nucl.\ Phys.\ B {\bf 31} (1971) 301.
  %%CITATION = NUPHA,B31,301;%%

%\cite{Slavnov:1972sq}
\bibitem{Slavnov:1972sq}
  A.~A.~Slavnov,
  %``Invariant regularization of gauge theories,''
  Theor.Math.Phys. {\bf 13} (1972) 1064
   [Teor.\ Mat.\ Fiz.\  {\bf 13} (1972) 174].
  %%CITATION = TMFZA,13,174;%%

%\cite{Krivoshchekov:1978xg}
\bibitem{Krivoshchekov:1978xg}
  V.~K.~Krivoshchekov,
  %``Invariant Regularizations for Supersymmetric Gauge Theories,''
  Theor.\ Math.\ Phys.\ {\bf 36} (1978) 745
 [Teor.\ Mat.\ Fiz.\  {\bf 36} (1978) 291].
 %%CITATION = TMFZA,36,291;%%

%\cite{West:1985jx}
\bibitem{West:1985jx}
  P.~C.~West,
  %``Higher Derivative Regulation Of Supersymmetric Theories,''
  Nucl.\ Phys.\ B {\bf 268} (1986) 113.
  %%CITATION = NUPHA,B268,113;%%

%\cite{Stepanyantz:2011jy}
\bibitem{Stepanyantz:2011jy}
  K.~V.~Stepanyantz,
  %``Derivation of the exact NSVZ $\beta$-function in N=1 SQED, regularized by higher derivatives, by direct summation of Feynman diagrams,''
  Nucl.\ Phys.\ B {\bf 852} (2011) 71.
  %%CITATION = ARXIV:1102.3772;%%

%\cite{Novikov:1983uc}
\bibitem{Novikov:1983uc}
  V.~A.~Novikov, M.~A.~Shifman, A.~I.~Vainshtein and V.~I.~Zakharov,
  %``Exact Gell-Mann-Low Function of Supersymmetric Yang-Mills Theories from Instanton Calculus,''
  Nucl.\ Phys.\ B {\bf 229} (1983) 381.
  %%CITATION = NUPHA,B229,381;%%

%\cite{Jones:1983ip}
\bibitem{Jones:1983ip}
  D.~R.~T.~Jones,
  %``More on the Axial Anomaly in Supersymmetric {Yang-Mills} Theory,''
  Phys.\ Lett.\ B {\bf 123} (1983) 45.
  %%CITATION = PHLTA,B123,45;%%

%\cite{Novikov:1985rd}
\bibitem{Novikov:1985rd}
  V.~A.~Novikov, M.~A.~Shifman, A.~I.~Vainshtein and V.~I.~Zakharov,
  %``Beta Function in Supersymmetric Gauge Theories: Instantons Versus Traditional Approach,''
  Phys.\ Lett.\ B {\bf 166} (1986) 329; Sov.\ J.\ Nucl.\ Phys.\  {\bf 43}
(1986) 294; [Yad.\ Fiz.\  {\bf 43} (1986) 459.]
  %%CITATION = PHLTA,B166,329;%%

%\cite{Shifman:1986zi}
\bibitem{Shifman:1986zi}
  M.~A.~Shifman and A.~I.~Vainshtein,
  %``Solution of the Anomaly Puzzle in SUSY Gauge Theories and the Wilson Operator Expansion,''
  Nucl.\ Phys.\ B {\bf 277}  (1986) 456; Sov.\ Phys.\ JETP {\bf 64} (1986) 428;
[Zh.\ Eksp.\ Teor.\ Fiz.\  {\bf 91}  (1986) 723.]
  %%CITATION = NUPHA,B277,456;%%

%\cite{Soloshenko:2003nc}
\bibitem{Soloshenko:2003nc}
  A.~A.~Soloshenko and K.~V.~Stepanyantz,
  %``Three loop beta function for N=1 supersymmetric electrodynamics, regularized by higher derivatives,''
  Theor.\ Math.\ Phys.\  {\bf 140} (2004) 1264
   [Teor.\ Mat.\ Fiz.\  {\bf 140} (2004) 430].
  %%CITATION = HEP-TH/0304083;%%

%\cite{Smilga:2004zr}
\bibitem{Smilga:2004zr}
  A.~V.~Smilga and A.~Vainshtein,
  %``Background field calculations and nonrenormalization theorems in 4-D supersymmetric gauge theories and their low-dimensional descendants,''
  Nucl.\ Phys.\ B {\bf 704} (2005) 445.
  %%CITATION = HEP-TH/0405142;%%

%\cite{Stepanyantz:2014ima}
\bibitem{Stepanyantz:2014ima}
  K.~V.~Stepanyantz,
  %``The NSVZ $\beta$-function and the Schwinger-Dyson equations for $\mathcal{N}=1$ SQED with $N_{f}$ flavors, regularized by higher derivatives,''
  JHEP {\bf 1408} (2014) 096.
  %[arXiv:1404.6717 [hep-th]].
  %%CITATION = ARXIV:1404.6717;%%

%\cite{Nartsev:2016nym}
\bibitem{Nartsev:2016nym}
  I.~V.~Nartsev and K.~V.~Stepanyantz,
  %``Exact renormalization of the photino mass in softly broken $ \mathcal{N} $ = 1 SQED with N$_{f}$ flavors regularized by higher derivatives,''
  JHEP {\bf 1704} (2017) 047.
  %doi:10.1007/JHEP04(2017)047
  %[arXiv:1610.01280 [hep-th]].
  %%CITATION = doi:10.1007/JHEP04(2017)047;%%

%\cite{Kazantsev:2014yna}
\bibitem{Kazantsev:2014yna}
  A.~E.~Kazantsev and K.~V.~Stepanyantz,
  %``Relation between two-point Green functions of ${\cal N}=1$ SQED with $N_f$ flavors, regularized by higher derivatives, in the three-loop approximation,''
  J.\ Exp.\ Theor.\ Phys. {\bf 120} (2015) 618 [Zh.\ Eksp.\ Teor.\ Fiz. {\bf 147} (2015) 714].
  %[arXiv:1410.1133 [hep-th]].
  %%CITATION = ARXIV:1410.1133;%%

%\cite{Aleshin:2015qqc}
\bibitem{Aleshin:2015qqc}
  S.~S.~Aleshin, A.~L.~Kataev and K.~V.~Stepanyantz,
  %``Structure of three-loop contributions to the beta-function of N=1 SQED with N_f flavors, regularized by the dimensional reduction,''
  Pisma Zh.\ Eksp.\ Teor.\ Fiz.\  {\bf 130} (2016) 83.
  %[arXiv:1511.05675 [hep-th]].
  %%CITATION = ARXIV:1511.05675;%%

%\cite{Aleshin:2016rrr}
\bibitem{Aleshin:2016rrr}
  S.~S.~Aleshin, I.~O.~Goriachuk, A.~L.~Kataev and K.~V.~Stepanyantz,
  %``The NSVZ scheme for ${\cal N}=1$ SQED with $N_f$ flavors, regularized by the dimensional reduction, in the three-loop approximation,''
  Phys.\ Lett.\ B {\bf 764} (2017) 222.
  %doi:10.1016/j.physletb.2016.11.041
  %[arXiv:1610.08034 [hep-th]].
  %%CITATION = doi:10.1016/j.physletb.2016.11.041;%%

%\cite{Vainshtein:1986ja}
\bibitem{Vainshtein:1986ja}
  A.~I.~Vainshtein, V.~I.~Zakharov and M.~A.~Shifman,
  %``Gell-mann-low Function In Supersymmetric Electrodynamics,''
  JETP Lett.\  {\bf 42} (1985) 224
 [Pisma Zh.\ Eksp.\ Teor.\ Fiz.\  {\bf 42} (1985) 182].
  %%CITATION = JTPLA,42,224;%%

%\cite{Shifman:1985fi}
\bibitem{Shifman:1985fi}
  M.~A.~Shifman, A.~I.~Vainshtein and V.~I.~Zakharov,
  %``Exact Gell-mann-low Function In Supersymmetric Electrodynamics,''
  Phys.\ Lett.\ B {\bf 166} (1986) 334.
  %%CITATION = PHLTA,B166,334;%%

%\cite{Shifman:1999mv}
\bibitem{Shifman:1999mv}
  M.~A.~Shifman and A.~I.~Vainshtein,
  %``Instantons versus supersymmetry: Fifteen years later,''
  In *Shifman, M.A.: ITEP lectures on particle physics and field theory, vol. 2* 485-647
  [hep-th/9902018].
  %%CITATION = HEP-TH/9902018;%%

%\cite{ArkaniHamed:1997mj}
\bibitem{ArkaniHamed:1997mj}
  N.~Arkani-Hamed and H.~Murayama,
  %``Holomorphy, rescaling anomalies and exact beta functions in supersymmetric gauge theories,''
  JHEP {\bf 0006} (2000) 030.
  %[hep-th/9707133].
  %%CITATION = HEP-TH/9707133;%%

%\cite{Kraus:2002nu}
\bibitem{Kraus:2002nu}
  E.~Kraus, C.~Rupp and K.~Sibold,
  %``Supersymmetric Yang-Mills theories with local coupling: The Supersymmetric gauge,''
  Nucl.\ Phys.\ B {\bf 661} (2003) 83.
  %[hep-th/0212064].
  %%CITATION = HEP-TH/0212064;%%

%\cite{Siegel:1979wq}
\bibitem{Siegel:1979wq}
  W.~Siegel,
  %``Supersymmetric Dimensional Regularization via Dimensional Reduction,''
  Phys.\ Lett.\ B {\bf 84} (1979) 193.
  %%CITATION = PHLTA,B84,193;%%

%\cite{Siegel:1980qs}
\bibitem{Siegel:1980qs}
  W.~Siegel,
  %``Inconsistency of Supersymmetric Dimensional Regularization,''
  Phys.\ Lett.\ B {\bf 94} (1980) 37.
  %%CITATION = PHLTA,B94,37;%%

%\cite{Avdeev:1981ew}
\bibitem{Avdeev:1981ew}
  L.~V.~Avdeev and O.~V.~Tarasov,
  %``The Three Loop Beta Function In The N=1, N=2, N=4 Supersymmetric Yang-mills Theories,''
  Phys.\ Lett.\ B {\bf 112} (1982) 356.
  %%CITATION = PHLTA,B112,356;%%

%\cite{Jack:1996vg}
\bibitem{Jack:1996vg}
  I.~Jack, D.~R.~T.~Jones and C.~G.~North,
  %``N=1 supersymmetry and the three loop gauge Beta function,''
  Phys.\ Lett.\ B {\bf 386} (1996) 138.

%\cite{Jack:1996cn}
\bibitem{Jack:1996cn}
  I.~Jack, D.~R.~T.~Jones and C.~G.~North,
  %``Scheme dependence and the NSVZ Beta function,''
  Nucl.\ Phys.\ B {\bf 486} (1997) 479.
  %%CITATION = HEP-PH/9609325;%%

%\cite{Harlander:2009mn}
\bibitem{Harlander:2009mn}
  R.~V.~Harlander, L.~Mihaila and M.~Steinhauser,
  %``The SUSY-QCD beta function to three loops,''
  Eur.\ Phys.\ J.\ C {\bf 63} (2009) 383.
  %doi:10.1140/epjc/s10052-009-1109-9
  %[arXiv:0905.4807 [hep-ph]].
  %%CITATION = doi:10.1140/epjc/s10052-009-1109-9;%%

%\cite{Harlander:2006xq}
\bibitem{Harlander:2006xq}
  R.~V.~Harlander, D.~R.~T.~Jones, P.~Kant, L.~Mihaila and M.~Steinhauser,
  %``Four-loop beta function and mass anomalous dimension in dimensional reduction,''
  JHEP {\bf 0612} (2006) 024.
  %%CITATION = HEP-PH/0610206;%%

%\cite{Jack:1998uj}
\bibitem{Jack:1998uj}
  I.~Jack, D.~R.~T.~Jones and A.~Pickering,
  %``The Connection between DRED and NSVZ,''
  Phys.\ Lett.\ B {\bf 435} (1998) 61.
  %%CITATION = HEP-PH/9805482;%%

%\cite{Kataev:2013csa}
\bibitem{Kataev:2013csa}
  A.~L.~Kataev and K.~V.~Stepanyantz,
  %``Scheme independent consequence of the NSVZ relation for $N=1$ SQED with $N_f$ flavors,''
  Phys.\ Lett.\ B {\bf 730} (2014) 184
  %[arXiv:1311.0589 [hep-th]].
  %%CITATION = ARXIV:1311.0589;%%

%\cite{Kataev:2014gxa}
\bibitem{Kataev:2014gxa}
  A.~L.~Kataev and K.~V.~Stepanyantz,
  %``The NSVZ $\beta$-function in supersymmetric theories with different regularizations and renormalization prescriptions,''
  Theor.\ Math.\ Phys.\  {\bf 181} (2014) 1531.
  %[arXiv:1405.7598 [hep-th]].
  %%CITATION = ARXIV:1405.7598;%%

%\cite{Kutasov:2004xu}
\bibitem{Kutasov:2004xu}
  D.~Kutasov and A.~Schwimmer,
  %``Lagrange multipliers and couplings in supersymmetric field theory,''
  Nucl.\ Phys.\ B {\bf 702} (2004) 369.
  %[hep-th/0409029].
  %%CITATION = HEP-TH/0409029;%%

%\cite{Avdeev:1981vf}
\bibitem{Avdeev:1981vf}
  L.~V.~Avdeev, G.~A.~Chochia and A.~A.~Vladimirov,
  %``On The Scope Of Supersymmetric Dimensional Regularization,''
  Phys.\ Lett.\ B {\bf 105} (1981) 272.
  %%CITATION = PHLTA,B105,272;%%

%\cite{Avdeev:1982xy}
\bibitem{Avdeev:1982xy}
  L.~V.~Avdeev and A.~A.~Vladimirov,
  %``Dimensional Regularization And Supersymmetry,''
  Nucl.\ Phys.\ B {\bf 219} (1983) 262.
  %%CITATION = NUPHA,B219,262;%%

%\cite{Pimenov:2009hv}
\bibitem{Pimenov:2009hv}
  A.~B.~Pimenov, E.~S.~Shevtsova and K.~V.~Stepanyantz,
  %``Calculation of two-loop beta-function for general N=1 supersymmetric Yang--Mills theory with the higher covariant derivative regularization,''
  Phys.\ Lett.\ B {\bf 686} (2010) 293.
  %%CITATION = ARXIV:0912.5191;%%

%\cite{Stepanyantz:2011bz}
\bibitem{Stepanyantz:2011bz}
  K.~V.~Stepanyantz,
  ``Factorization of integrals defining the two-loop $\beta$-function for the general renormalizable N=1 SYM theory, regularized by the higher covariant derivatives, into integrals of double total derivatives,''
  arXiv:1108.1491 [hep-th].
  %%CITATION = ARXIV:1108.1491;%%

%\cite{Aleshin:2016yvj}
\bibitem{Aleshin:2016yvj}
  S.~S.~Aleshin, A.~E.~Kazantsev, M.~B.~Skoptsov and K.~V.~Stepanyantz,
  %``One-loop divergences in non-Abelian supersymmetric theories regularized by BRST-invariant version of the higher derivative regularization,''
  JHEP {\bf 1605} (2016) 014.
  %doi:10.1007/JHEP05(2016)014
  %[arXiv:1603.04347 [hep-th]].
  %%CITATION = doi:10.1007/JHEP05(2016)014;%%

%\cite{Shakhmanov:2017soc}
\bibitem{Shakhmanov:2017soc}
  V.~Y.~Shakhmanov and K.~V.~Stepanyantz,
  %``Three-loop NSVZ relation for terms quartic in the Yukawa couplings with the higher covariant derivative regularization,''
  Nucl.\ Phys.\ B {\bf 920} (2017) 345.
  %doi:10.1016/j.nuclphysb.2017.04.017
  %[arXiv:1703.10569 [hep-th]].
  %%CITATION = doi:10.1016/j.nuclphysb.2017.04.017;%%

%\cite{Kataev:2013eta}
\bibitem{Kataev:2013eta}
  A.~L.~Kataev and K.~V.~Stepanyantz,
  %``NSVZ scheme with the higher derivative regularization for $\mathcal{N} =$ 1 SQED,''
  Nucl.\ Phys.\ B {\bf 875} (2013) 459.
  %%CITATION = ARXIV:1305.7094;%%

%\cite{Stepanyantz:2016gtk}
\bibitem{Stepanyantz:2016gtk}
  K.~V.~Stepanyantz,
  %``Non-renormalization of the $V\bar cc$-vertices in ${\cal N}=1$ supersymmetric theories,''
  Nucl.\ Phys.\ B {\bf 909} (2016) 316.
  %doi:10.1016/j.nuclphysb.2016.05.011
  %[arXiv:1603.04801 [hep-th]].
  %%CITATION = doi:10.1016/j.nuclphysb.2016.05.011;%%

%\cite{Hisano:1997ua}
\bibitem{Hisano:1997ua}
  J.~Hisano and M.~A.~Shifman,
  %``Exact results for soft supersymmetry breaking parameters in supersymmetric gauge theories,''
  Phys.\ Rev.\ D {\bf 56} (1997) 5475.
  %doi:10.1103/PhysRevD.56.5475
  %[hep-ph/9705417].
  %%CITATION = doi:10.1103/PhysRevD.56.5475;%%

%\cite{Jack:1997pa}
\bibitem{Jack:1997pa}
  I.~Jack and D.~R.~T.~Jones,
  %``The Gaugino Beta function,''
  Phys.\ Lett.\ B {\bf 415} (1997) 383.
  %doi:10.1016/S0370-2693(97)01277-X
  %[hep-ph/9709364].
  %%CITATION = doi:10.1016/S0370-2693(97)01277-X;%%

%\cite{Avdeev:1997vx}
\bibitem{Avdeev:1997vx}
  L.~V.~Avdeev, D.~I.~Kazakov and I.~N.~Kondrashuk,
  %``Renormalizations in softly broken SUSY gauge theories,''
  Nucl.\ Phys.\ B {\bf 510} (1998) 289.
  %doi:10.1016/S0550-3213(98)81015-8, 10.1016/S0550-3213(97)00706-2
  %[hep-ph/9709397].
  %%CITATION = doi:10.1016/S0550-3213(98)81015-8, 10.1016/S0550-3213(97)00706-2;%%

%\cite{Nartsev:2016mvn}
\bibitem{Nartsev:2016mvn}
  I.~V.~Nartsev and K.~V.~Stepanyantz,
  %``NSVZ-like scheme for the photino mass in softly broken ${\cal N}=1$ SQED regularized by higher derivatives,''
  JETP Lett.\  {\bf 105} (2017) no.2,  69.
  %doi:10.1134/S0021364017020059
  %[arXiv:1611.09091 [hep-th]].
  %%CITATION = doi:10.1134/S0021364017020059;%%

%\cite{Slavnov:2001pu}
\bibitem{Slavnov:2001pu}
  A.~A.~Slavnov,
  %``Universal gauge invariant renormalization,''
  Phys.\ Lett.\ B {\bf 518} (2001) 195.
  %doi:10.1016/S0370-2693(01)01002-4
  %%CITATION = doi:10.1016/S0370-2693(01)01002-4;%%

%\cite{Slavnov:2002ir}
\bibitem{Slavnov:2002ir}
  A.~A.~Slavnov,
  %``Regularization-independent gauge-invariant renormalization of the Yang-Mills theory,''
  Theor.\ Math.\ Phys.\  {\bf 130} (2002) 1
   [Teor.\ Mat.\ Fiz.\  {\bf 130} (2002) 3].
  %doi:10.1023/A:1013828529525
  %%CITATION = doi:10.1023/A:1013828529525;%%

%\cite{Slavnov:2002kg}
\bibitem{Slavnov:2002kg}
  A.~A.~Slavnov and K.~V.~Stepanyantz,
  %``Universal invariant renormalization for supersymmetric theories,''
  Theor.\ Math.\ Phys.\  {\bf 135} (2003) 673
   [Teor.\ Mat.\ Fiz.\  {\bf 135} (2003) 265].
  %doi:10.1023/A:1023622616220
  %[hep-th/0208006].
  %%CITATION = doi:10.1023/A:1023622616220;%%

%\cite{Slavnov:2003cx}
\bibitem{Slavnov:2003cx}
  A.~A.~Slavnov and K.~V.~Stepanyantz,
  %``Universal invariant renormalization of supersymmetric Yang-Mills theory,''
  Theor.\ Math.\ Phys.\  {\bf 139} (2004) 599
   [Teor.\ Mat.\ Fiz.\  {\bf 139} (2004) 179].
  %doi:10.1023/B:TAMP.0000026178.67671.6a
  %[hep-th/0305128].
  %%CITATION = doi:10.1023/B:TAMP.0000026178.67671.6a;%%

%\cite{Becchi:1974md}
\bibitem{Becchi:1974md}
  C.~Becchi, A.~Rouet and R.~Stora,
  %``Renormalization of the Abelian Higgs-Kibble Model,''
  Commun.\ Math.\ Phys.\  {\bf 42} (1975) 127.
  %%CITATION = CMPHA,42,127;%%

%\cite{Tyutin:1975qk}
\bibitem{Tyutin:1975qk}
  I.~V.~Tyutin,
  %``Gauge Invariance in Field Theory and Statistical Physics in Operator Formalism,''
  Lebedev Institute preprint No. 39 (1975), arXiv:0812.0580 [hep-th].
  %%CITATION = ARXIV:0812.0580;%%

%\cite{Piguet:1981fb}
\bibitem{Piguet:1981fb}
  O.~Piguet and K.~Sibold,
  %``Renormalization of $N=1$ Supersymmetrical {Yang-Mills} Theories. 1. The Classical Theory,''
  Nucl.\ Phys.\ B {\bf 197} (1982) 257.
  doi:10.1016/0550-3213(82)90291-7
  %%CITATION = doi:10.1016/0550-3213(82)90291-7;%%

%\cite{Faddeev:1980be}
\bibitem{Faddeev:1980be}
  L.~D.~Faddeev and A.~A.~Slavnov,
``Gauge Fields. Introduction To Quantum Theory,'' Nauka, Moscow,
1978 and
  Front.\ Phys.\  {\bf 50} (1980) 1
   [Front.\ Phys.\  {\bf 83} (1990) 1].
  %%CITATION = FRPHA,50,1;%%

%\cite{Slavnov:1977zf}
\bibitem{Slavnov:1977zf}
  A.~A.~Slavnov,
  %``The Pauli-Villars Regularization for Nonabelian Gauge Theories,''
  Theor.\ Math.\ Phys.\ {\bf 33} (1977) 977
   [Teor.\ Mat.\ Fiz.\  {\bf 33} (1977) 210].
  %%CITATION = TMFZA,33,210;%%

%\cite{Taylor:1971ff}
\bibitem{Taylor:1971ff}
  J.~C.~Taylor,
  %``Ward Identities and Charge Renormalization of the Yang-Mills Field,''
  Nucl.\ Phys.\ B {\bf 33} (1971) 436.
  %%CITATION = NUPHA,B33,436;%%

%\cite{Slavnov:1972fg}
\bibitem{Slavnov:1972fg}
  A.~A.~Slavnov,
  %``Ward Identities in Gauge Theories,''
  Theor.\ Math.\ Phys.\  {\bf 10} (1972) 99
   [Teor.\ Mat.\ Fiz.\  {\bf 10} (1972) 153].
  %%CITATION = TMPHA,10,99;%%

%\cite{Kazantsev:2017fdc}
\bibitem{Kazantsev:2017fdc}
  A.~E.~Kazantsev, M.~B.~Skoptsov and K.~V.~Stepanyantz,
  ``One-loop polarization operator of the quantum gauge superfield for ${\cal N}=1$ SYM regularized by higher derivatives,''
  arXiv:1709.08575 [hep-th].
  %%CITATION = ARXIV:1709.08575;%%

%\cite{Bogolyubov:1980nc}
\bibitem{Bogolyubov:1980nc}
  N.~N.~Bogolyubov and D.~V.~Shirkov,
  ``Introduction To The Theory Of Quantized Fields,''
  Nauka, Moscow, 1984
  [Intersci.\ Monogr.\ Phys.\ Astron.\  {\bf 3} (1959) 1].
  %%CITATION = IMTPA,3,1;%%

%\cite{Collins:1984xc}
\bibitem{Collins:1984xc}
  J.~C.~Collins,
  ``Renormalization : An Introduction to Renormalization, The Renormalization Group, and the Operator Product Expansion.''
  %%CITATION = INSPIRE-209810;%%

%\cite{Broadhurst:1999he}
\bibitem{Broadhurst:1999he}
  D.~J.~Broadhurst,
  %``Four loop Dyson-Schwinger-Johnson anatomy,''
  Phys.\ Lett.\ B {\bf 466} (1999) 319.
  %doi:10.1016/S0370-2693(99)01083-7
  %[hep-ph/9909336].
  %%CITATION = doi:10.1016/S0370-2693(99)01083-7;%%

%\cite{Kataev:2010tm}
\bibitem{Kataev:2010tm}
  A.~L.~Kataev,
  %``Riemann $\zeta(3)$- terms in perturbative QED series, conformal symmetry and the analogies with structures of multiloop effects in N=4 supersymmetric Yang-Mills theory,''
  Phys.\ Lett.\ B {\bf 691} (2010) 82.
  %doi:10.1016/j.physletb.2010.06.005
  %[arXiv:1005.2058 [hep-th]].
  %%CITATION = doi:10.1016/j.physletb.2010.06.005;%%

%\cite{Kataev:2013vua}
\bibitem{Kataev:2013vua}
  A.~L.~Kataev,
  %``Conformal symmetry limit of QED and QCD and identities between perturbative contributions to deep-inelastic scattering sum rules,''
  JHEP {\bf 1402} (2014) 092.
  %doi:10.1007/JHEP02(2014)092
  %[arXiv:1305.4605 [hep-th]].
  %%CITATION = doi:10.1007/JHEP02(2014)092;%%

%\cite{Soloshenko:2002np}
\bibitem{Soloshenko:2002np}
  A.~Soloshenko and K.~Stepanyantz,
  ``Two loop renormalization of N=1 supersymmetric electrodynamics, regularized by higher derivatives,''
  hep-th/0203118.
  %%CITATION = HEP-TH/0203118;%%

%\cite{Soloshenko:2003sx}
\bibitem{Soloshenko:2003sx}
  A.~A.~Soloshenko and K.~V.~Stepanyantz,
  %``Two-loop anomalous dimension of N = 1 supersymmetric quantum electrodynamics regularized using higher covariant derivatives,''
  Theor.\ Math.\ Phys.\  {\bf 134} (2003) 377
   [Teor.\ Mat.\ Fiz.\  {\bf 134} (2003) 430].
  %doi:10.1023/A:1022653506397
  %%CITATION = doi:10.1023/A:1022653506397;%%

\end{thebibliography}
\end{document}